%% file: Marulli.tex
\documentclass[longauth]{aa}

\usepackage{graphicx}
\usepackage{txfonts}
\usepackage{subfigure}
\usepackage{float}
\usepackage{natbib,twoopt}
\usepackage{xspace}
\usepackage{balance}
\usepackage{upgreek}
\usepackage{multirow}
\usepackage[colorlinks=true]{hyperref}

\def\apjl{Astrophys. J. Lett.}
\def\apjs{Astrophys. J. Suppl.}

\def\aap{Astron. Astrophys.}

\newcommand{\Mpch}{$h^{-1}\,\mbox{Mpc}$\,}
\newcommand{\Mpchc}{$h^{-3}\,\mbox{Mpc}^{3}$\,}
\newcommand{\xiiz}{$\xi(r_p,\pi)$\,}

\bibpunct{(}{)}{;}{a}{}{,} 
\DeclareGraphicsExtensions{.eps,.ps,.pdf}

\begin{document}

\title{The VIMOS Public Extragalactic Redshift Survey (VIPERS)
  \thanks{Based on observations collected at the European Southern
    Observatory, Paranal, Chile, under programmes 182.A-0886 (LP) at
    the Very Large Telescope, and also based on observations obtained
    with MegaPrime/MegaCam, a joint project of CFHT and CEA/DAPNIA, at
    the Canada-France-Hawaii Telescope (CFHT), which is operated by
    the National Research Council (NRC) of Canada, the Institut
    National des Science de l'Univers of the Centre National de la
    Recherche Scientifique (CNRS) of France, and the University of
    Hawaii. This work is based in part on data products produced at
    TERAPIX and the Canadian Astronomy Data Centre as part of the
    Canada-France-Hawaii Telescope Legacy Survey, a collaborative
    project of NRC and CNRS. The VIPERS web site is
    http://vipers.inaf.it/.}  }

\subtitle{Luminosity and stellar mass dependence of galaxy clustering
  at $0.5<z<1.1$}

\include{authors}

\date{Received --; accepted --}

\abstract 
{}
{We investigate the dependence of galaxy clustering on luminosity and
  stellar mass in the redshift range $0.5<z<1.1$, using the first
  $\sim55000$ redshifts from the VIMOS Public Extragalactic Redshift
  Survey (VIPERS).}
{We measured the redshift-space two-point correlation functions
  (2PCF), $\xi(s)$ and \xiiz, and the projected correlation function,
  $w_p(r_p)$, in samples covering different ranges of B-band absolute
  magnitudes and stellar masses. We considered both threshold and
  binned galaxy samples, with median B-band absolute magnitudes
  $-21.6\lesssim M_{\rm B}-5\log(h)\lesssim-19.5$ and median stellar
  masses $9.8\lesssim\log(M_\star[h^{-2}\,M_\odot])\lesssim10.7$. We
  assessed the real-space clustering in the data from the projected
  correlation function, which we model as a power law in the range
  $0.2<r_p[$\Mpch$]<20$. Finally, we estimated the galaxy bias as a
  function of luminosity, stellar mass, and redshift, assuming a flat
  $\Lambda$ cold dark matter model to derive the dark matter 2PCF.}
{We provide the best-fit parameters of the power-law model assumed for
  the real-space 2PCF -- the correlation length, $r_0$, and the slope,
  $\gamma$ -- as well as the linear bias parameter, as a function of
  the B-band absolute magnitude, stellar mass, and redshift. We
  confirm and provide the tightest constraints on the dependence of
  clustering on luminosity at $0.5<z<1.1$. We prove the complexity of
  comparing the clustering dependence on stellar mass from samples
  that are originally flux-limited and discuss the possible origin of
  the observed discrepancies.  Overall, our measurements provide
  stronger constraints on galaxy formation models, which are now
  required to match, in addition to local observations, the clustering
  evolution measured by VIPERS galaxies between $z=0.5$ and $z=1.1$
  for a broad range of luminosities and stellar masses.}
{}

\keywords{Cosmology: observations -- Cosmology: large-scale
      structure of Universe -- Surveys -- Galaxies: evolution }
    
\authorrunning{F. Marulli et al.}
\titlerunning{Clustering as a function of luminosity and stellar mass}

\maketitle

\section{Introduction} 

The large scale structure (LSS) of the Universe contains fundamental
information for investigating the nature of the dark matter (DM) and
the origin of the accelerated expansion of the Universe. However,
since the LSS is traced by galaxies, the relevant information can only
be extracted after specifying the mapping between galaxies and DM
\citep{kaiser1984}. The way galaxies trace the underlying density
field, i.e. their biasing, can be understood through the mean
occupation of galaxies in DM haloes, under the assumption that all DM
mass in the Universe is in the form of haloes. This relation depends
on the galaxy properties and the sample selection.

Indeed, it has been observationally established in the local Universe
that the spatial distribution of galaxies significantly depends on
their luminosity, stellar mass, colour, morphological, and spectral
types, as well as on the environment \citep[see e.g.][]{davis1976,
  giovanelli1986, davis1988, hamilton1988, einasto1991,
  maurogordato1991, loveday1995, benoist1996, guzzo1997, guzzo2000,
  willmer1998, brown2000, brown2003, norberg2001, norberg2002,
  zehavi2002, zehavi2005, madgwick2003, abbas2006, zehavi2011,
  budavari2003, li2006, swanson2008, loh2010, ross2011, guoh2013,
  christodoulou2012}.  The general trend is that luminous and massive
galaxies with bulge-dominated morphologies and red colours cluster
more strongly than faint, less massive, blue, spiral galaxies, an
effect that becomes particularly evident above the characteristic
luminosity, $L^\star$, and stellar mass, $M^\star$, of the Schechter
function. This is expected in the standard cosmological scenario of
hierarchical growth of cosmic structures, where bright galaxies are
hosted by massive, rare, and highly clustered DM haloes
\citep{kaiser1984, white1987}.

Observations up to intermediate redshifts, $z\sim1$, i.e. half way
through the cosmic time from the microwave background to the present,
show clustering trends similar to those observed in the local
Universe, with the more luminous and massive galaxies being more
clustered than the faint, less massive ones \citep[see
  e.g.][]{coil2006, coil2008, mccracken2008, meneux2006, meneux2008,
  meneux2009, pollo2006, abbas2010, coupon2012, mostek2013}.  At
higher redshifts, $z>1$, observational constraints are generally
weaker, suffering from a combination of many types of selection
biases, but the measurements still show a luminosity and stellar mass
segregation in galaxy clustering \citep[see e.g.][]{daddi2003,
  adelberger2005b, ouchi2005, lee2006, lee2009, hildebrandt2009,
  wake2011, lin2012}, with strong evidence of a steepening in the
correlation function at $r\lesssim 1$\Mpch with increasing luminosity.

It is possible to derive cosmological constraints from the spatial
properties of galaxies using methods that do not explicitly depend on
galaxy bias, such as by exploiting baryonic acoustic oscillations
\citep[see e.g.][]{eisenstein2005,cole2005,percival2007,percival2010}
or geometric clustering distortions \citep[see e.g.][]{alcock1979,
  simpson2010, marulli2012b}, or by combining the one- and two-point
moments of the smoothed galaxy density distribution
\citep{bel2013}. However, to extract the full information encoded in
the shape of the galaxy power spectrum and in its redshift-space
distortions, the galaxy bias must be accounted for to some
extent. Since the connection between baryons and DM can be
significantly affected by non-linear phenomena, such as mergers,
dynamical friction, cooling, feedback, etc., semi-analytic models or
full hydrodynamic simulations provide useful tools for properly
investigating how galaxies trace the DM distribution. Once a
mathematical description of the galaxy bias has been obtained,
cosmological constraints can be derived by marginalizing over the
uncertain parameters that describe the bias. Alternatively,
observational measurements of the galaxy bias, derived in a given
cosmological framework, can be used to constrain the parameters of
galaxy formation models.

In this paper, we focus on the dependence of galaxy clustering on
luminosity and stellar mass, providing new observational constraints
at redshift $0.5<z<1.1$, from the first data release of the ongoing
VIMOS Public Extragalactic Redshift Survey
\citep[VIPERS,][]{guzzo2013}. The comparison between VIPERS clustering
measurements with similar ones at lower and higher redshifts can
constrain the cosmic evolution of the relationship between DM and
galaxy properties, hence between gravity and cosmology on one side and
processes associated with baryonic physics on the other side. The
large number of objects with accurate spectroscopic redshifts and the
large volume covered by the VIPERS survey give us the opportunity to
measure the clustering of galaxies over a wide range of luminosities
and stellar masses, back to the cosmic epoch when the Universe was
still decelerating. In particular, we provide constraints on the
projected correlation function, which we model as a single power
law. This represents a reasonable, though not optimal, approximation
at $r_p\lesssim 20$\Mpch, where $r_p$ is the direction perpendicular
to the line-of-sight, and facilitates the comparison with previous
studies. A more detailed interpretation of these data using halo
occupation distribution (HOD) modelling is deferred to a future work
\citep[see also][]{delatorre2013b}.

The plan of this paper is as follows.  In Section \ref{sec:data} we
give a general overview of the VIPERS data, describing in detail the
luminosity and stellar mass selections adopted in this work. We
investigate the spatial distribution of VIPERS galaxies measuring the
two-point correlation function (2PCF) in approximately volume-limited
samples of different B-band absolute magnitudes and stellar
masses. The methodology of this analysis is described in Section
\ref{sec:method}, where we also explain how we account for the
incompleteness affecting our VIPERS samples. More technical details
about the method to correct for the proximity effect and stellar mass
incompleteness are provided in Appendix~\ref{appendix}. The results of
our analysis are presented in Section \ref{sec:results}. Finally, in
Section \ref{sec:concl} we summarize the main results of this work.

Throughout this paper, we assume a flat $\Lambda$ cold dark matter
(CDM) background cosmology, with mass density parameter $\Omega_{\rm
  M}=0.25$, both to convert the measured redshifts into comoving
distances and to estimate the DM correlation used to assess the galaxy
bias. Changing the value of $\Omega_{\rm M}$ in a reasonable range,
given the current cosmological constraints, would only introduce small
effects on our clustering measurements within their observational
uncertainties \citep[see e.g.][]{marulli2012b}.  Galaxy magnitudes are
quoted in the AB system. The dependence of observed quantities on the
Hubble parameter is indicated as a function of $h\equiv H_0/100\, {\rm
  km\, s^{-1} Mpc^{-1}}$.

\begin{figure*}
  \includegraphics[width=0.49\textwidth]{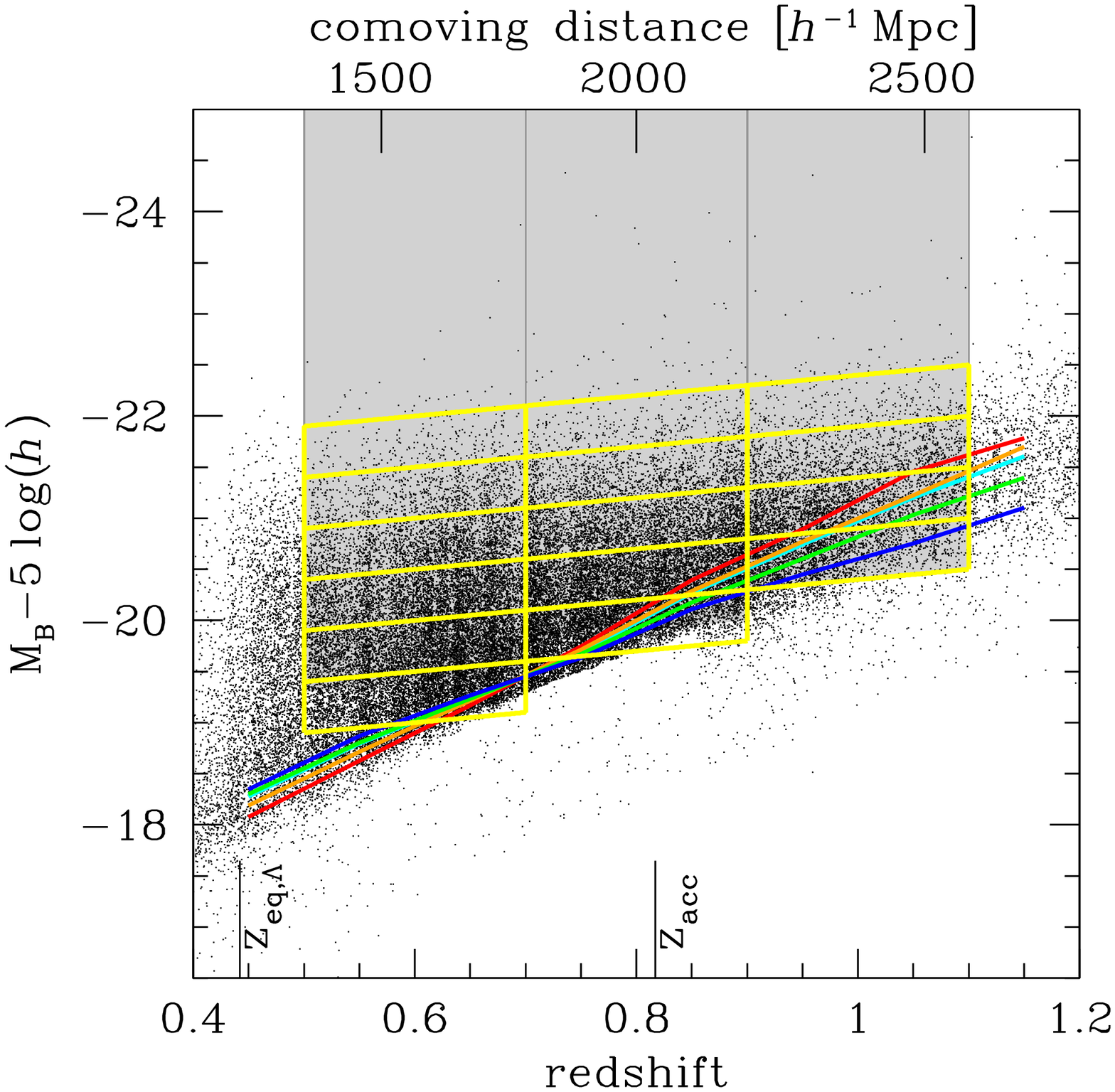}
  \includegraphics[width=0.49\textwidth]{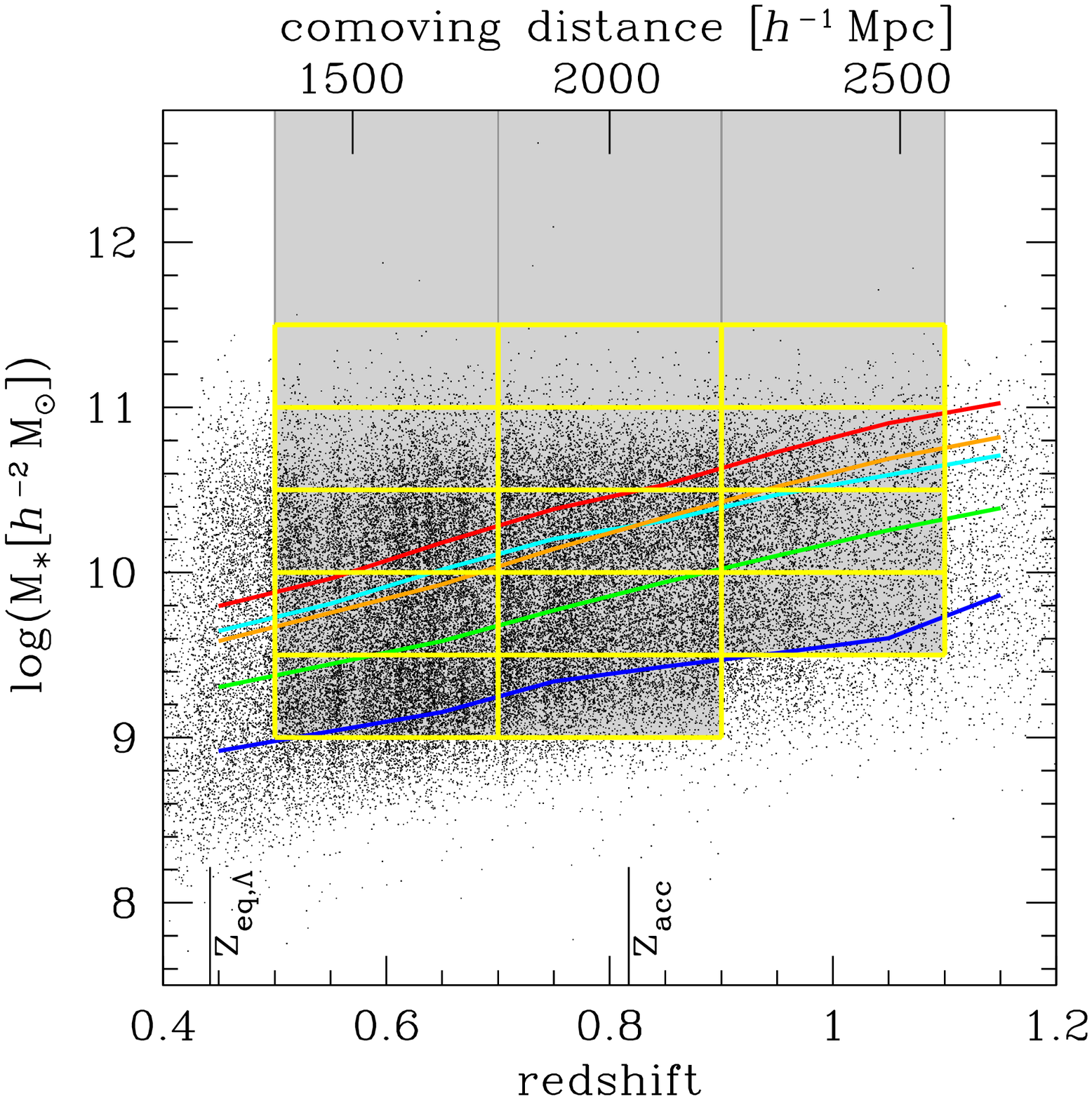}
  \caption{Luminosity and stellar mass selection as a function of
    redshift. The black dots show the W1 and W4 VIPERS galaxies.
    Yellow lines surround the selected sub-samples. Solid lines show
    the $90$\% completeness levels for the whole galaxy population
    (cyan lines) and for galaxies of four different photometric types:
    red E/Sa (red lines), early spiral (orange lines), late spiral
    (green lines), and irregular or starburst (blue lines). The B-band
    rest-frame magnitude limits in the left panel account for the
    average luminosity evolution of galaxies, as explained in the
    text. On the bottom axes, {\large $z_{\rm eq,\Lambda}$}$\sim0.44$
    is the redshift of DM-DE equality and {\large $z_{\rm
        acc}$}$\sim0.82$ is the redshift when the Universe started
    accelerating, according to $\Lambda $CDM predictions with the set
    of cosmological parameters adopted in this work.}
  \label{fig:sub_samples}
\end{figure*}


\section{The data}
\label{sec:data}
\subsection{The VIPERS survey}

We have used the data from the VIPERS Public Data Release 1 (PDR-1),
which will be made publicly available in the fall of 2013. VIPERS is an
ongoing Large Programme aimed at measuring redshifts for $\sim 10^{5}$
galaxies at $0.5 < z \lesssim 1.2$. One of the main goals of this
survey is to characterize the galaxy spatial properties and growth
rate of the LSS, measuring galaxy clustering and redshift-space
distortions at an epoch when the Universe was about half its current
age. Indeed, according to the standard $\Lambda $CDM scenario, the
redshift range that VIPERS probes is previous to the dark energy (DE)
dominated era, crossing the cosmic epoch when the Universe started
accelerating (see the bottom labels in Fig.~\ref{fig:sub_samples}).

The galaxy target sample is selected from the Canada-France-Hawaii
Telescope Legacy Survey Wide (CFHTLS-Wide) optical photometric
catalogues \citep{mellier2008}.  VIPERS covers $\sim24$ deg$^2$ on the
sky, divided over two areas within the W1 and W4 CFHTLS fields.
Galaxies are selected down to a flux limit of $i_{AB}<22.5$, and a
robust $gri$ colour pre-selection is applied to effectively remove
objects at $z<0.5$. Coupled with an aggressive observing strategy
\citep{scodeggio2009}, this allows us to double the galaxy sampling
rate in the redshift range of interest ($\sim 40$\%), with respect to
a pure magnitude-limited sample.  At the same time, the area and depth
of the survey result in a fairly large volume, $5 \times 10^{7}$
\Mpchc, analogous to that of the 2dFGRS at $z\sim0.1$.  Such a
combination of sampling and depth is unique among current redshift
surveys at $z>0.5$.  VIPERS is performed with the VIMOS multi-object
spectrograph at the European Southern Observatory Very Large Telescope
\citep{lefevre2002, lefevre2003}, at moderate resolution ($R=210$),
using the LR Red grism, providing a wavelength coverage of
5500-9500$\rm{\AA}$ and a typical radial velocity error of 140 km
sec$^{-1}$ .  The full VIPERS area is covered through a mosaic of 288
VIMOS pointings (192 in the W1 area and 96 in the W4 area).  A
complete description of the VIPERS survey, from the definition of the
target sample to the actual spectra and redshift measurements, is
presented in the parallel survey description paper \citep{guzzo2013},
while a discussion of the survey data reduction and management
infrastructure can be found in \citet{garilli2012}.

The PDR-1 galaxy catalogue includes $53,608$ redshifts and corresponds
to the reduced data frozen in the VIPERS database at the end of the
2011/2012 observing campaign.  The 2PCF of PDR-1 galaxies and the
first constraints on the growth rate from redshift-space distortions
are presented in a parallel paper \citep{delatorre2013b}. A
measurement of $\Omega_{\rm M}$ using a novel clustering statistic is
given in \citet{bel2013}.  The galaxy stellar mass function is
presented in \citet{davidzon2013}, while the luminosity function and
the evolution of the colour bimodality is provided in
\citet{fritz2013}.  The PDR-1 galaxies are classified with a support
vector machine algorithm in \citet{malek2013}, while an early subset
of the spectra has been analysed and classified through a principal
component analysis in \citet{marchetti2013}. Finally, the VIPERS
redshift distribution has been used to measure the real-space galaxy
power spectrum with colour-selected samples from the CFHTLS
\citep{granett2012}.


\subsection{Luminosity and mass selection}

We consider galaxy samples selected in three redshift ranges:
$z\in[0.5,0.7]$, $z\in[0.7,0.9]$ and $z\in[0.9,1.1]$. To investigate
the luminosity and mass dependence of galaxy clustering, we use both
threshold and binned sub-samples with different selections in B-band
absolute magnitudes, $M_{\rm B}$, and stellar masses, $M_{\star}$. The
latter quantities have been estimated with a spectral energy
distribution (SED) fitting technique, using the {\small HYPERZMASS}
code \citep{bolzonella2000, bolzonella2010}.  Absolute magnitudes are
measured using the B Buser filter in the AB system.  More details
about the SED fitting procedure are given in \citet{davidzon2013}.

Figure \ref{fig:sub_samples} shows the sub-samples selected in this
work. The B-band magnitude limits account for the average luminosity
evolution of galaxies. Following \citet{meneux2009}, we model the
average redshift evolution of galaxy magnitudes as $M_{\rm
  B}(z)=M_{\rm B}(0)-z$ \citep{ilbert2005, zucca2009}. Instead, for
the stellar mass-selected sub-samples, we adopt the more conservative
and widely used flat thresholds, since the redshift evolution of
$M_\star$ is negligible up to $z\sim1$ \citep{pozzetti2007,
  pozzetti2010, davidzon2013}. The solid coloured lines represent the
$90$\% completeness levels for the whole galaxy population and for
galaxies of four different photometric types: red E/Sa, early spiral,
late spiral and irregular or starburst \citep[see][for more
  details]{davidzon2013}. We did not use these completeness limits to
investigate the dependence of galaxy clustering on spectral type,
which we defer to a future work. These lines are displayed here to
show that our magnitude-selected samples are all volume-limited
irrespective of their colours, with the exception of the last redshift
bin, where our $i$ band selection makes us start losing the reddest
galaxy population. However, the mass-selected samples are affected by
the incompleteness of red galaxies. We will come back to this point in
Section \ref{sub:proxy_mass} and Appendix~\ref{appendix}.

Finally, in the bottom axes of Fig.~\ref{fig:sub_samples}, we mark
{\large $z_{\rm eq,\Lambda}$}, the redshift of DM-DE equality, and
{\large $z_{\rm acc}$}, the redshift when the Universe started
accelerating, according to $\Lambda $CDM predictions with the set of
cosmological parameters adopted in this work. Clearly, the three
redshift ranges considered for the present analysis probe different
cosmic epochs, crossing the transition from deceleration to
acceleration.


\section{The two-point correlation function}
\label{sec:method}

In this section, we describe the methodology used to measure the
galaxy 2PCF. In particular, we discuss our error estimates and
sampling rate correction.

\subsection{Clustering measurements}

We quantify the spatial properties of the VIPERS galaxies using the
2PCF, $\xi(r)$. This function is implicitly defined as
$dP=n^2[1+\xi(r)]dV_1dV_2$, where $dP$ is the probability of finding a
pair of galaxies in the comoving volumes $dV_1$ and $dV_2$, separated
by a comoving distance $r$, and $n$ is the galaxy average number
density. In the following, we refer to real-space and redshift-space
three-dimensional separations using the vectors $\vec{r}$ and
$\vec{s}$, respectively.

To measure the function $\xi(s)$, we make use of the standard
\citet{landy1993} estimator:
\begin{equation}
\xi(s)=\frac{GG(s)-2GR(s)+RR(s)}{RR(s)} \, ,
\label{eq:landy}
\end{equation}
where $GG(s)$, $GR(s)$, and $RR(s)$ are the normalized galaxy--galaxy,
galaxy--random, and random--random pairs, with spatial separation in
the range $[s-\Delta s/2, s+\Delta s/2]$, respectively.

\begin{figure}
  \includegraphics[width=0.49\textwidth]{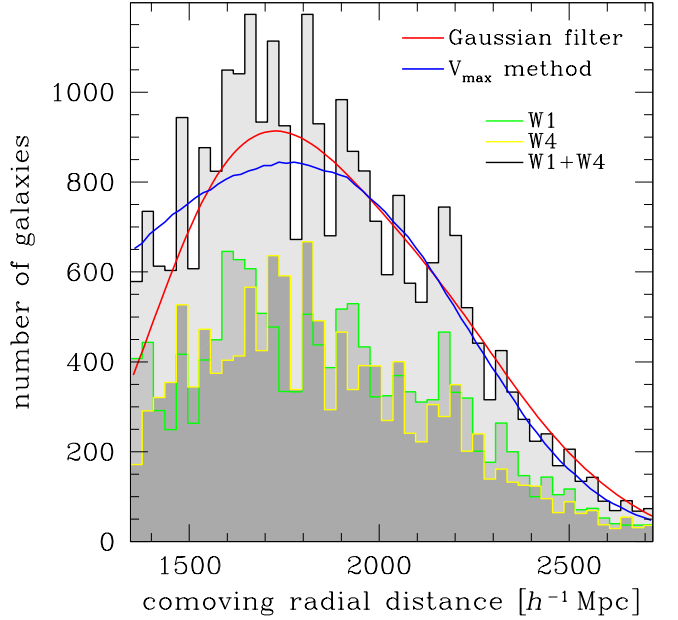}
  \caption{Comoving radial distance distribution of the W1 and W4
    VIPERS galaxy samples, shown with the green and yellow histograms,
    respectively. The red and blue lines show the total W1+W4 smoothed
    distribution, obtained by filtering the observed distribution with
    a Gaussian kernel of $\sigma=150$\Mpch and using the $V_{\rm max}$
    method, respectively.}
  \label{fig:dc_distr}
\end{figure}

The random samples are generated in the right ascension-declination
plane, adopting the same geometric mask of the real data. We set
$N_R=30N_{gal}$, where $N_R$ is the number of random objects and
$N_{gal}$ the number of galaxies in each sample. The latter set of
values is reported in the fifth columns of Tables
~\ref{tab:table1}--\ref{tab:table4}.  The redshifts of the random
objects are extracted from the smoothed radial distribution of the
W1+W4 VIPERS galaxies, estimated in each luminosity and stellar mass
sub-sample. Figure \ref{fig:dc_distr} shows the whole comoving radial
distances distribution of the W1 and W4 VIPERS galaxies. To smooth the
total W1+W4 radial distribution, we use the non-parametric $V_{\rm
  max}$ method \citep[see][for more details]{delatorre2013b}. A
similar result can be obtained by filtering the distribution with a
Gaussian kernel of $\sigma=150$\Mpch. The advantage of the $V_{\rm
  max}$ method is that it does not require us to make assumptions
about the kernel.  We have verified that all the results presented in
this paper do not vary significantly when changing the smoothing
scheme adopted.

\subsection{Error estimate}
\label{sub:errors}

The errors on the clustering measurements are estimated using a set of
mock galaxy catalogues. For the luminosity-selected samples, we use
data-calibrated HOD mocks, while for the stellar mass-selected
samples, we consider galaxy mocks built from the stellar-to-halo mass
relation (SHMR) of \citet{moster2013}. We use 26 independent HOD and
SHMR mock catalogues, both constructed by assigning galaxies to DM
haloes of the MultiDark simulation, a large N-body run in the $\Lambda
$CDM framework \citep{prada2012}. A novel technique has been used to
repopulate the MultiDark simulation with haloes of mass below the
resolution limit \citep{delatorre2013a} and thus to be able to
simulate the faintest and less massive galaxies observed in VIPERS. A
detailed description of the method used to construct our mock galaxy
catalogues can be found in \citet{delatorre2013b}. The covariance
error matrix is estimated from the dispersion among the mock
catalogues:
\begin{equation}
  C_{ij}=\frac{1}{N-1}\sum_{k=1}^{N}\left(\bar{\xi}_i -
  \xi_i^k\right)\left(\bar{\xi}_j-\xi_j^k\right) \, ,
\label{eq:covariance}
\end{equation}
where $\bar{\xi}_i$ is the mean value of the 2PCF, $\xi_i^k$, measured
in each mock catalogue.  The errors on the 2PCF are then obtained from
the square root of the diagonal values of the covariance matrix,
$\sigma_i=\sqrt{C_{ii}}$, which include both the Poissonian noise and
the uncertainties due to the size of the volume explored, i.e. the
sample variance. The luminosity and stellar mass dependence of the
2PCF errors is estimated by measuring the covariance matrix in mock
sub-samples with the same luminosity and stellar mass selection used
for the real data.


\subsection{Weighting the galaxy pairs}
\label{sub:ww}

To properly account for the effects of sampling rate and colour
selection, each galaxy has been weighted according to its sky position
and redshift, using the following weight function:
\begin{eqnarray}
w(Q,z) & = & w_{CSR}\cdot w_{TSR}\cdot w_{SSR} \nonumber \\ & = &
CSR^{-1}(z)\cdot TSR^{-1}(Q)\cdot SSR^{-1}(Q) \, .
\label{eq:weight}
\end{eqnarray}
All of the weights in Eq.~(\ref{eq:weight}) are derived directly from
the data. The colour sampling rate, $CSR$, represents the completeness
due to the colour selection. It mainly depends on redshift and can be
modelled as $CSR(z)=0.5[1-{\rm erf}(7.405-17.465\cdot z)]$, where
${\rm erf}$ is the error function ${\rm
  erf}(x)=2/\sqrt\pi\int_0^x\exp(-t^2)dt$ \citep{guzzo2013,
  coupon2013}. The target sampling rate, $TSR$, is defined as the
fraction of galaxies in the photometric catalogue that have been
spectroscopically targeted, $TSR=N_{\rm gal}(0\leq z_{\rm
  flag}\leq9.5)/N_{\rm target}$, where $z_{\rm flag}$ is the quality
flag of the spectroscopic redshift measurement (see
\citealt{guzzo2013} for more details on $z_{\rm flag}$). In addition,
we have a $3.2$\% stellar contamination in the target sample, but the
effect on the $TSR$ is negligible.  Finally, the spectroscopic success
rate, $SSR$, is defined as the fraction of the spectroscopically
targeted galaxies for which a secure identification (i.e. high-quality
flag) has been obtained, $SSR=N_{\rm gal}(2\leq z_{\rm
  flag}\leq9.5)/N_{\rm gal}(0\leq z_{\rm flag}\leq9.5)$. Both the
$TSR$ and $SSR$ depend on the VIMOS quadrant, $Q$, in which the galaxy
has been observed. All the other possible dependencies, such as on
redshift and magnitude, are comparatively smaller and can be
neglected, as discussed in detail in \citet{delatorre2013b}.

\begin{figure*}
  \includegraphics[width=\textwidth]{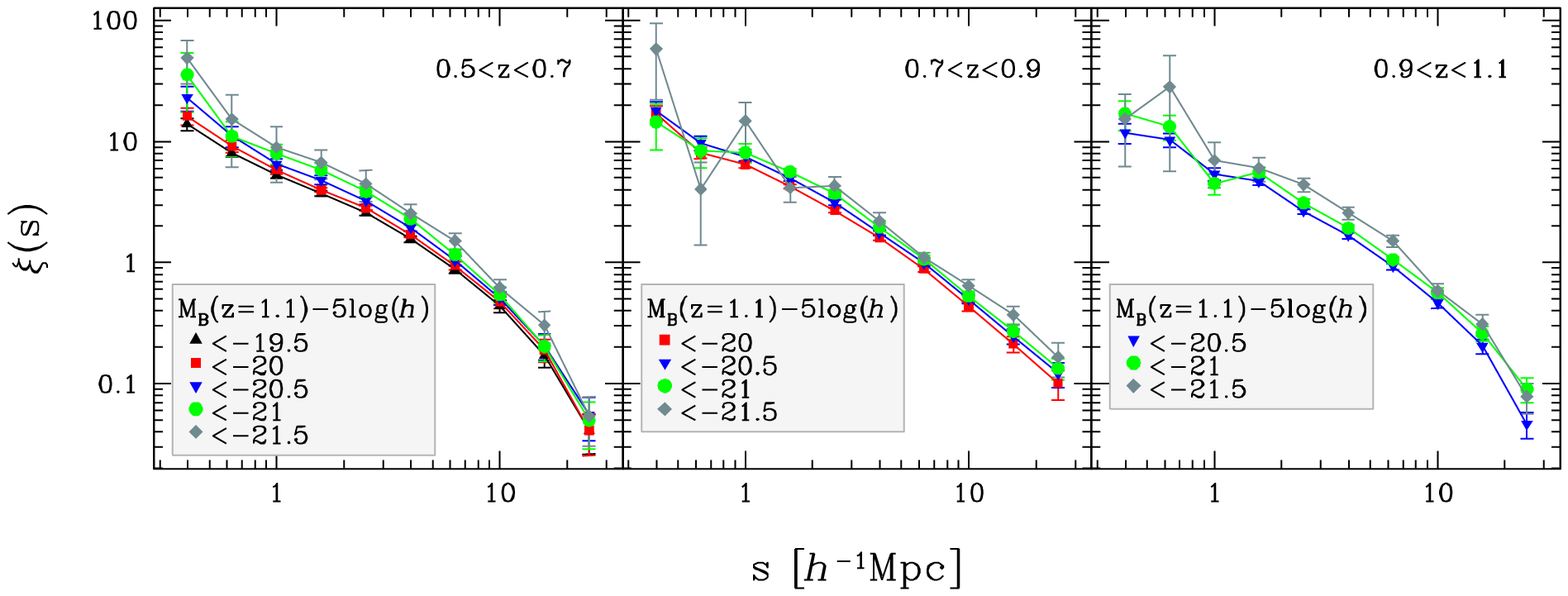}
  \includegraphics[width=\textwidth]{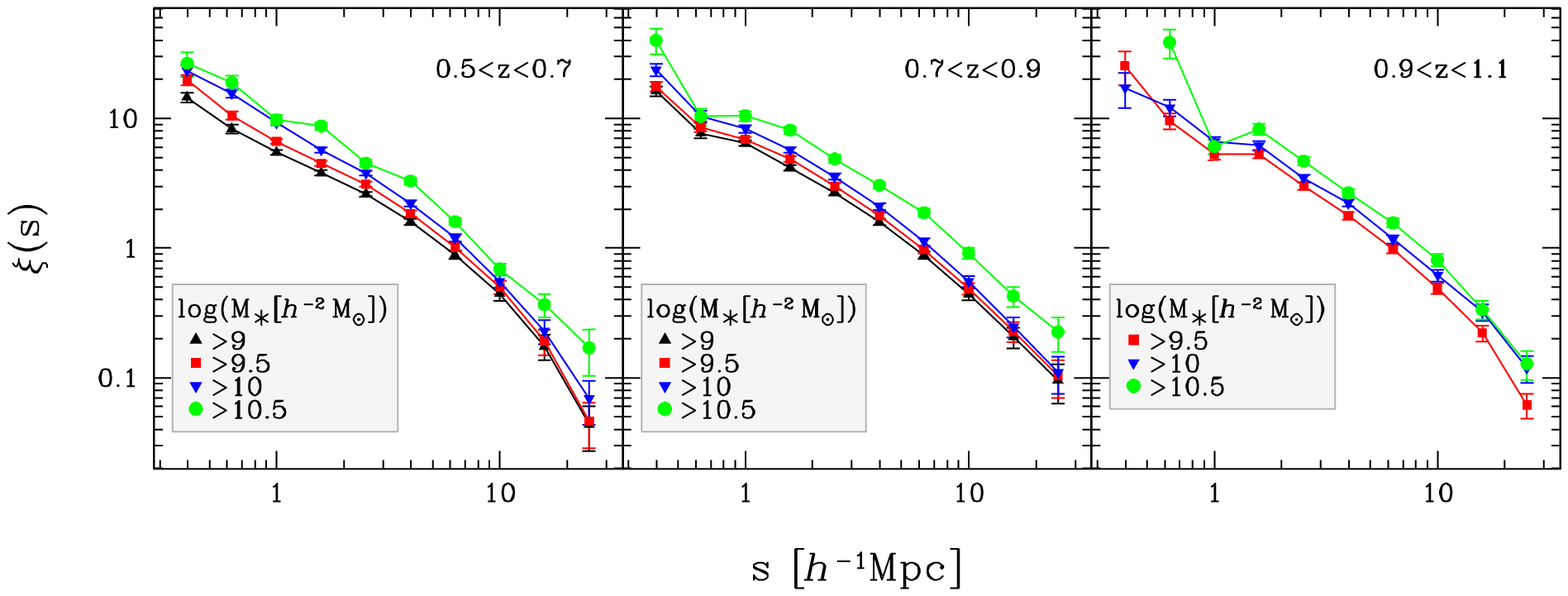}
  \caption{2PCF of VIPERS galaxies as a function of B-band absolute
    magnitude ({\em upper panels}) and stellar mass ({\em bottom
      panels}). The error bars are the square root of the diagonal
    values of the covariance matrix given by Eq.~\ref{eq:covariance}.
    Absolute magnitudes are given in the AB system.}
  \label{fig:xi_zspace}
\end{figure*}

\begin{figure*}
  \includegraphics[width=\textwidth]{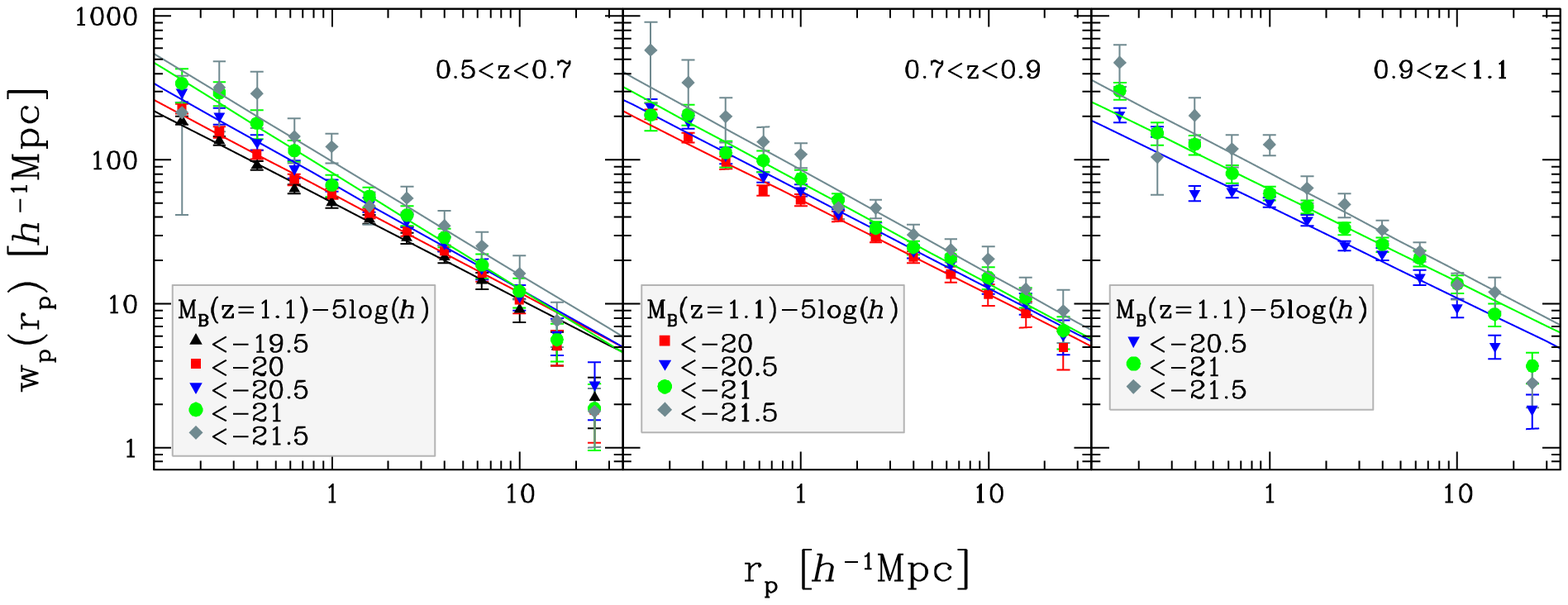}
  \includegraphics[width=\textwidth]{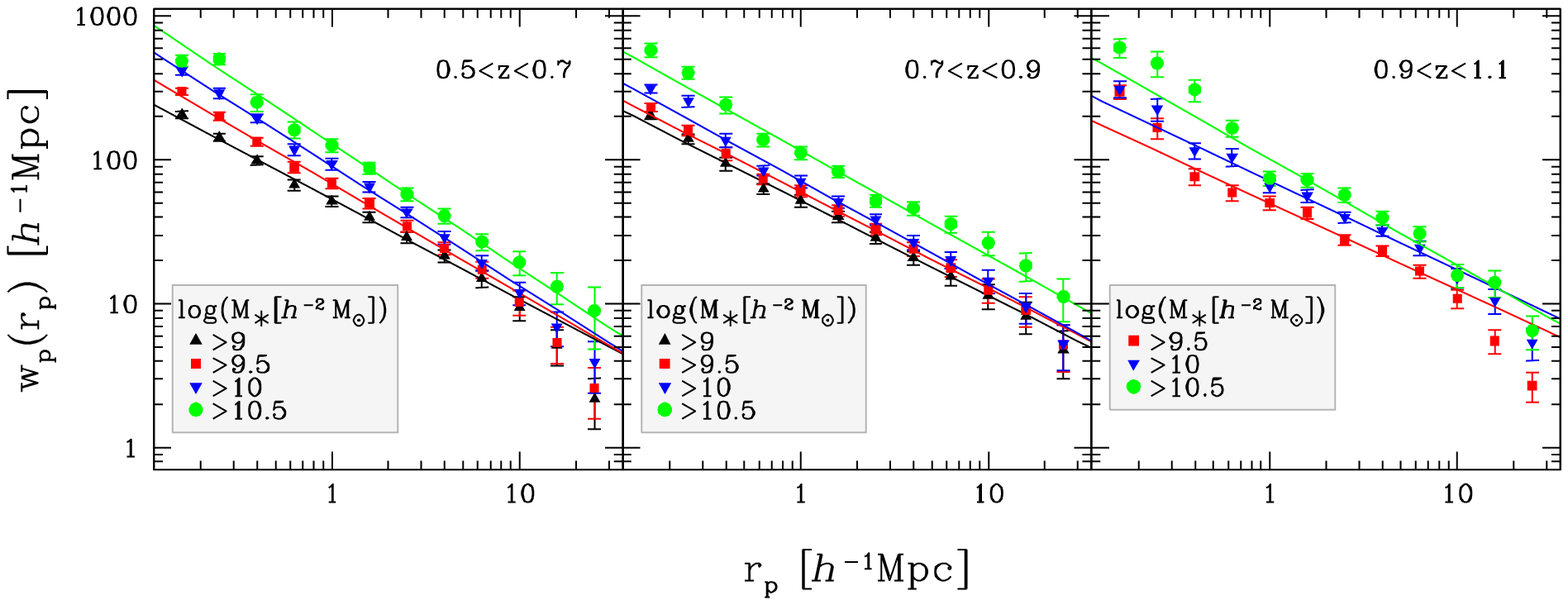}
  \caption{Projected correlation function of VIPERS galaxies as a
    function of B-band absolute magnitude ({\em upper panels}) and
    stellar mass ({\em bottom panels}). The solid lines show the
    best-fit power-law models, obtained by fitting the correlation
    functions in the interval $0.2<r_p[$\Mpch$]<20$. }
  \label{fig:xi_proj}
\end{figure*}


\subsection {Proximity effects and mass incompleteness}
\label{sub:proxy_mass}

The VIMOS multi-object spectrograph allows spectra for $\sim$400
galaxies to be obtained in a single pointing \citep{bottini2005};
however, it is not possible to target galaxies that are close to each
other along the dispersion direction.  Thus, the highest density
regions are under-sampled, resulting in a systematic under-estimation
of the correlation function, especially on small scales
\citep{pollo2005, delatorre2013b}. To estimate the impact of the
proximity effect, we measured the mean ratio between the
redshift-space 2PCF in two sets of mock catalogues with and without
the slit mask target selection algorithm applied. A similar test has
been applied by \citet{coil2008}.  The mock catalogues used here are
constructed with a semi-analytic model \citep{delucia2007} on top of
the Millennium simulation \citep{springel2005}. To obtain a smoothed
version of this ratio, we modelled it with an error function (see
Fig.~\ref{fig:SSPOC} in Appendix~\ref{appendix}). Then, we multiplied
the measured redshift-space 2PCF by this smoothed ratio.  Since we
find this correction to be almost independent of luminosity and
stellar mass, we used the values obtained in our largest galaxy sample
to correct all of our clustering measurements. At $z\lesssim0.9$, the
under-estimation of the 2PCF due to the proximity effect is $\sim10$\%
on scales $r\gtrsim1$ \Mpch, and increases slightly at smaller
separations.

As shown in the right-hand panel of Fig.~\ref{fig:sub_samples}, our
mass-selected galaxy samples are partially incomplete due to the
limiting $i_{AB}<22.5$ flux cut of VIPERS, missing galaxies with high
mass-to-light ratio. To correct for this effect, we would have to know
what are the clustering properties of the galaxies that are missing in
each of our samples. This can only be inferred using galaxy formation
models both complete in mass and realistically describing the
formation and evolution of faint galaxies.

Following what has already been done in previous studies \citep[see
  e.g.][]{meneux2008, meneux2009}, we implemented a correction method
based on a strategy similar to the one described above for the
proximity effect. Specifically, the correction is derived by measuring
the ratio between the mean redshift-space 2PCF of two sets of mock
catalogues, one complete in mass (down to $\sim10^8 M_\odot$) and the
other one magnitude-limited at $i_{AB}<22.5$. According to the
semi-analytic model considered here, the high mass-to-light ratio
galaxies that are missed in our samples are faint red objects, which
are predicted to be more clustered than the average population. As a
result, the stellar mass incompleteness is expected to introduce a
scale-dependent clustering suppression, up to $\sim$50\% on scales
$\lesssim 1$ \Mpch, for the redshift-space 2PCF.

This correction assumes that different types of mock galaxies have
clustering properties that match those of the real objects.  However,
this is only partially true.  It is known that the model considered
here overpredicts the abundance of faint and red galaxies.  In
addition, while the clustering of blue galaxies is well reproduced,
the mock red galaxies appear more clustered than in real data
\citep{delatorre2011, cucciati2012}. For the above reasons, such a
model-dependent correction should be applied with caution. In the
following, we show our clustering measurements without correcting for
stellar mass incompleteness. Nevertheless, in
Tables~\ref{tab:table3}--\ref{tab:table4}, we provide our results
obtained both with and without this correction, to directly show the
expected effect according to the semi-analytic model considered.  More
details about these two corrections for the proximity effect and mass
incompleteness are provided in Appendix~\ref{appendix}.

\begin{figure*}
  \includegraphics[width=\textwidth]{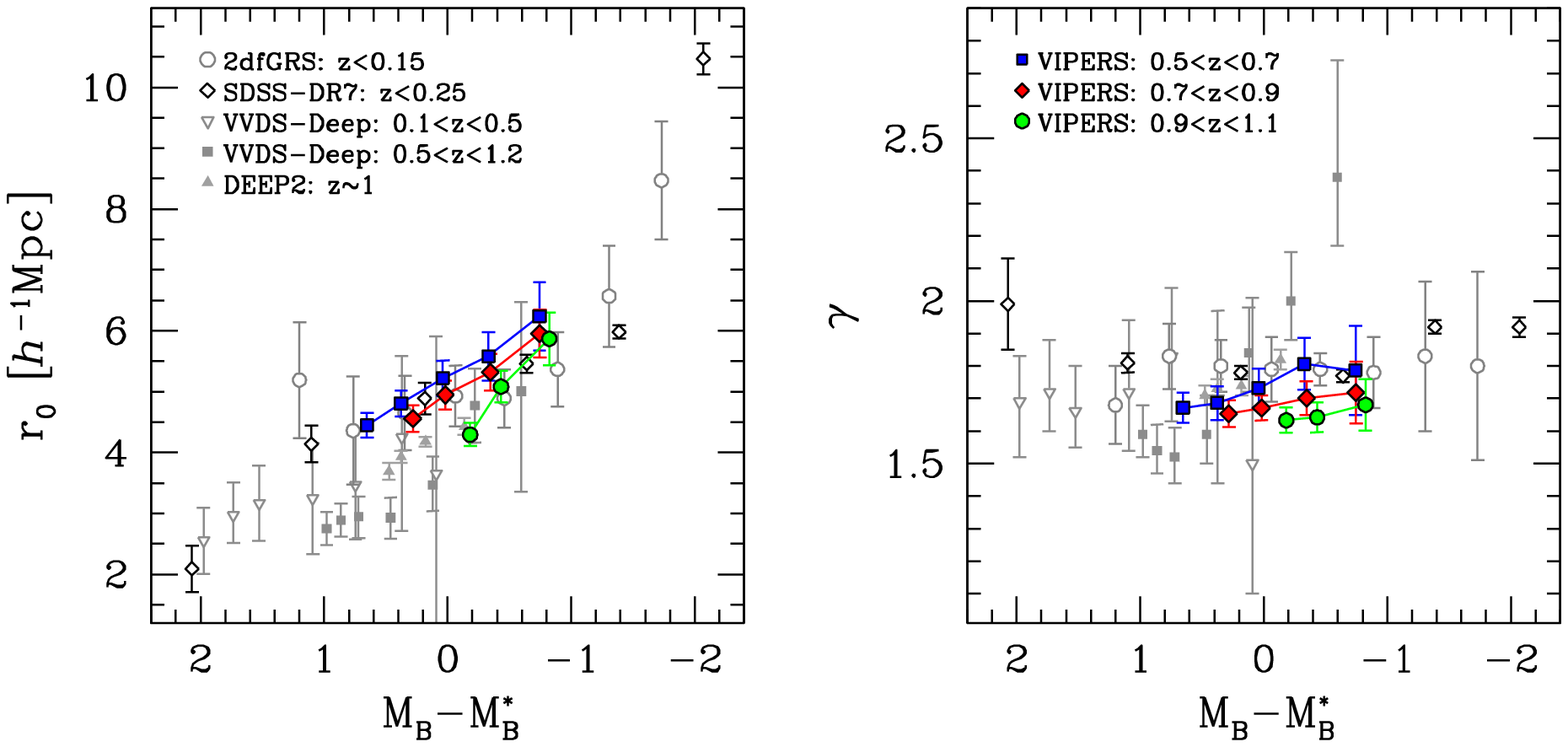}
  \includegraphics[width=\textwidth]{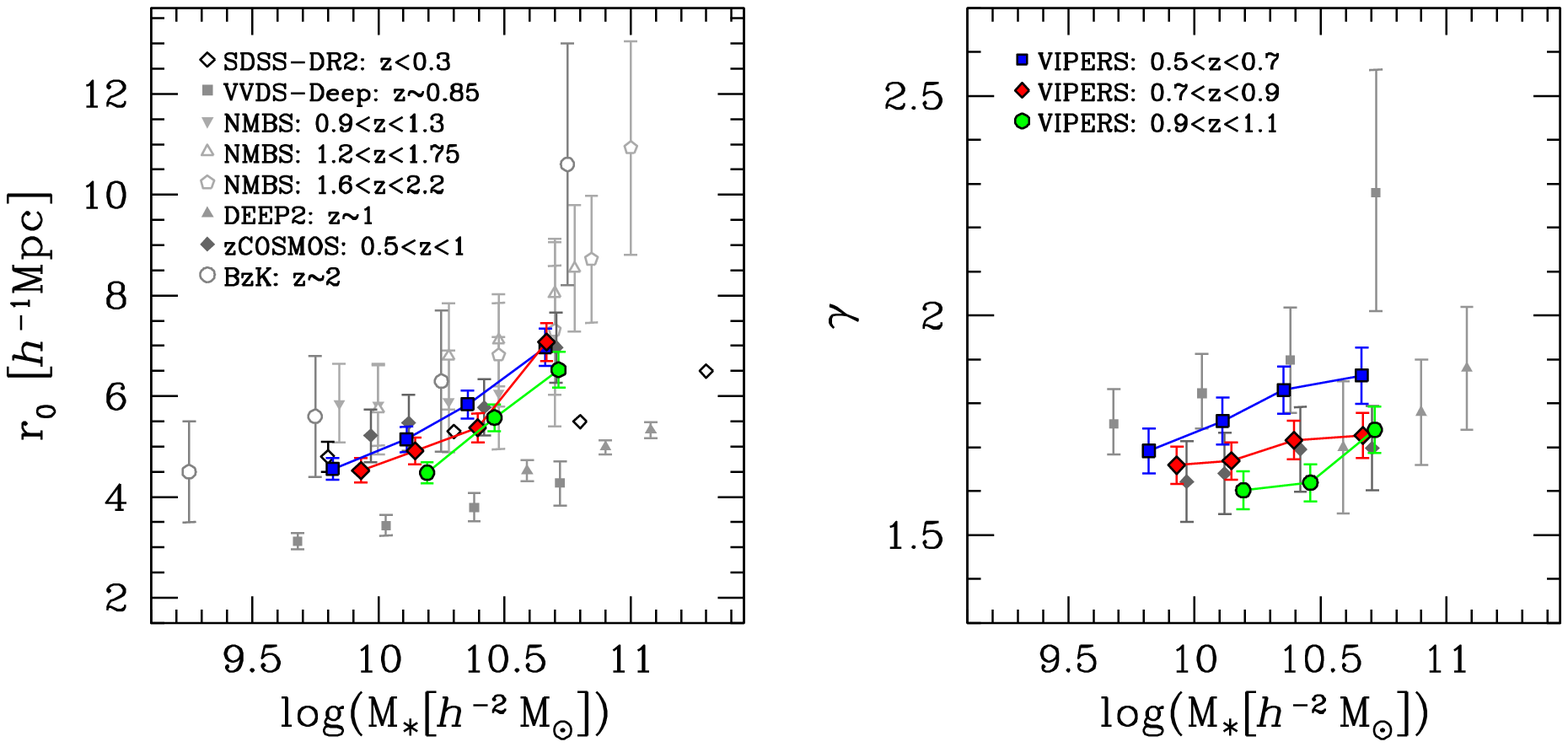}
  \caption{Best-fit values of the correlation length, $r_0$, ({\em
      left panels}) and slope, $\gamma$, ({\em right panels}), as a
    function of B-band absolute magnitude ({\em upper panels}) and
    stellar mass ({\em bottom panels}). VIPERS costraints, obtained in
    threshold samples at $z\in[0.5,0.7]$, $z\in[0.7,0.9]$, and
    $z\in[0.9,1.1]$ are shown, respectively, by blue squares, red
    diamonds, and green circles, as indicated by the labels. The
    remaining grey symbols refer to measurements based on data from
    other surveys, reported here for comparison: {\em upper panels} --
    DSS-DR7 data from \citet{zehavi2011}, 2dfGRS data from
    \citet{norberg2002}, VVDS-Deep data from \citet{pollo2006} and
    DEEP2 data from \citet{coil2006}; {\em lower panels} -- SDSS-DR2
    data from \citet{li2006}, VVDS-Deep data from \citet{meneux2008},
    zCOSMOS data from \citet{meneux2009}, NMBS data from
    \citet{wake2011}, DEEP2 data from \citet{mostek2013} and BzK data
    from \citet{lin2012}.}
\label{fig:r0_gamma}
\end{figure*}


\subsection{Projected and real-space correlations and galaxy bias}

To correct for redshift-space distortions due to galaxy peculiar
velocities and to derive the 2PCF in real space, $\xi(r)$, we first
measure the clustering as a function of the two distances
perpendicular and parallel to the line-of-sight, \xiiz, using the
\citet{landy1993} estimator.  Then, we integrate along the line of
sight, $\pi$, as
\begin{equation} 
  w_p(r_p) = 2\int^{\pi_{\rm max}}_{0} d\pi' \,  \xi(r_p, \pi')
  \, .
\label{eq:xi_proj}
\end{equation} 
The real-space 2PCF can then be estimated by {\em deprojecting} the
function $w_p(r_p)$ \citep{davis1983, saunders1992}:
\begin{equation} 
\xi(r) = -\frac{1}{\pi}\int^{\infty}_r dr_p'
\frac{dw_p(r_p')/dr_p}{\sqrt{r_p'^2-r^2}} \; .
\label{eq:xideproj} 
\end{equation}
To facilitate the comparison with previous studies, we do not use
Eq.~\ref{eq:xideproj}, but model the function $\xi(r)$ as a single
power law:
\begin{equation} 
\xi(r)=\left(\frac{r}{r_0}\right)^{-\gamma} \, .
\label{eq:powerlaw}
\end{equation} 
With the above assumption, Eq.~(\ref{eq:xi_proj}) can be solved
analytically:
\begin{equation}
  w_p(r_p)=r_p\left(\frac{r_0}{r_p}\right)^\gamma
  \frac{\Gamma(\frac{1}{2})\Gamma(\frac{\gamma-1}{2})}{\Gamma(\frac{\gamma}{2})} \, ,
\label{eq:xi_proj_analit}
\end{equation}
where $\Gamma$ is the Euler's Gamma function.  Hence, a power-law fit
to $w_p(r_p)$ provides $r_0$ and $\gamma$ for the real-space
correlation function $\xi(r)$.  Finally, the scale-dependent galaxy
bias can be estimated as
\begin{equation} 
b(r_p)=\sqrt{\frac{w_p(r_p)}{w_p^{\rm m}(r_p)}} \, ,
\label{eq:bias}
\end{equation}
where $w_p^{\rm m}(r_p)$ is the projected correlation function of
matter.  This is derived using the software {\small CAMB}
\citep{lewis2002}, which exploits the {\small HALOFIT} routine
\citep{smith2003} to reproduce the non-linear evolution of the matter
power spectrum.


\section{Results}
\label{sec:results}

In this section, we present our results for the luminosity and
stellar-mass dependence of galaxy clustering and galaxy bias, and
compare them with those of previous studies.

\subsection{Redshift-space and projected 2PCF}

Figure \ref{fig:xi_zspace} shows the redshift-space 2PCF of VIPERS
galaxies, $\xi(s)$, as a function of B-band absolute magnitude ({\em
  upper panels}) and stellar mass ({\em bottom panels}), in the three
redshift ranges $z\in[0.5,0.7]$, $z\in[0.7,0.9]$, and
$z\in[0.9,1.1]$. The function $\xi(s)$ has been estimated using
Eq.~(\ref{eq:landy}), while the error bars are the square root of the
diagonal values of the covariance matrix, given by
Eq.~(\ref{eq:covariance}).

To derive the real-space 2PCF and directly compare the VIPERS
clustering with previous data, we estimate the projected correlation
function using Eq.~\ref{eq:xi_proj}. We integrate \xiiz up to
$\pi_{\rm max}=30$ \Mpch. As we verified, this choice represents a
convenient compromise between robustness and the need to exclude noisy
bins at large separations. The projected 2PCF of VIPERS galaxies is
shown in Fig.~\ref{fig:xi_proj}. The upper and lower panels show,
respectively, the luminosity and stellar mass dependence of
$w_p(r_p)$, as indicated by the labels. The best-fit power-law models,
given by Eq.~\ref{eq:xi_proj_analit}, have been obtained by fitting
the correlation functions in the interval $0.2<r_p[$\Mpch$]<20$. We
can see in Fig.~\ref{fig:xi_proj} that the power-law model does not
describe the small-scale behaviour of $w_p(r_p)$ well for the most
massive galaxies.  This steepening in the amplitude of the correlation
function on scales below $1$\Mpch can be interpreted in the framework
of the halo model as due to the larger fraction of satellite galaxies
in the massive haloes, creating a prominent one-halo term in the
correlation function. Nevertheless, the results presented in this work
are robust for the scale range used for the fit, given the estimated
uncertainties.

We find a clear luminosity and stellar mass segregation in galaxy
clustering at all redshifts considered. This result agrees with
previous findings, as discussed in detail in the next section, and
provides new strong constraints on galaxy formation models in the
intermediate redshift Universe.  The clustering measurements presented
in Figs. \ref{fig:xi_zspace} and \ref{fig:xi_proj} refer to
threshold galaxy samples, as indicated by the labels.  Our
measurements on the galaxy clustering in both threshold and binned
galaxy samples are given in
Tables~\ref{tab:table1}--\ref{tab:table4}. Specifically, we provide
the best-fit parameters of the power-law model: the correlation length
$r_0$ (sixth columns), and the slope $\gamma$ (seventh columns).  The
errors on the best-fit parameters are derived from the scatter of the
$r_0$ and $\gamma$ values among the 26 independent VIPERS mocks.

\begin{figure}
  \includegraphics[width=0.49\textwidth]{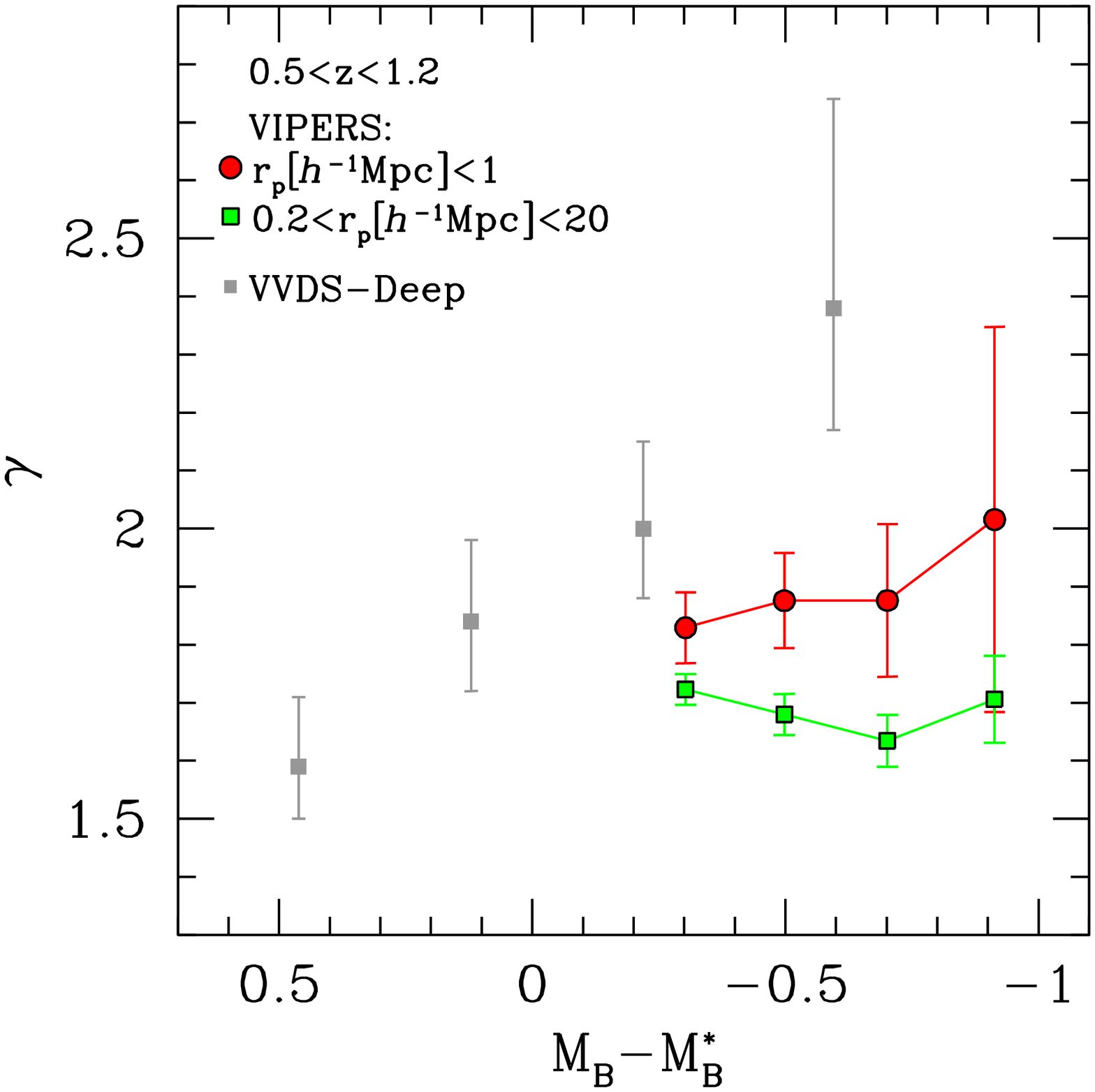}
  \caption{Best-fit values of the VIPERS correlation slope, $\gamma$,
    as a function of luminosity. Green and red points show the VIPERS
    $\gamma$ values obtained fitting $w_p(r_p)$ in
    $0.2<r_p[$\Mpch$]<20$ and $r_p<2$ \Mpch, respectively. Grey
    squares refer to the VVDS-Deep measurements.}
  \label{fig:gamma_MagB}
\end{figure}

\subsection{Comparison to previous studies}
\label{sec:comparison}

In Fig.~\ref{fig:r0_gamma}, we compare our VIPERS measurements with
previous results, at both low and high redshift. The upper panels show
the best-fit values of $r_0$ and $\gamma$ as a function of the B-band
magnitude. Following \citet{pollo2006}, we normalize the median
magnitudes of each galaxy sample to the corresponding values of the
characteristic magnitude of the Schechter luminosity function, $M_{\rm
  B}^*$, obtained by fitting the values of $M_{\rm B}^*(z)$ given by
\citet{ilbert2005}.  With this method, we can properly compare samples
at different redshifts, taking the mean brightening of galaxies due to
their evolution into account.  A clear luminosity evolution of the
clustering length, $r_0$, is evident at all redshifts. We can see that
$r_0$ increases with time between $z\sim1$ and $z\sim0.6$. Moreover,
there is some evidence that $r_0$ grows faster for faint galaxies
compared to the bright ones.

For the local Universe, we consider the data from \citet{zehavi2011}
and \citet{norberg2002}. \citet{zehavi2011} measured the luminosity
dependence of galaxy clustering using the Sloan Digital Sky Survey
Seventh Data Release (SDSS-DR7), with volume-limited galaxy samples up
to $z=0.25$. \citet{norberg2002} used the 2dF Galaxy Redshift Survey
(2dFGRS) to measure the galaxy clustering up to $z=0.15$. The
VIMOS-VLT Deep Survey (VVDS) clustering measurements by
\citep{pollo2006} refer to the redshift ranges $0.1<z<0.5$ and
$0.5<z<1.2$ . Finally, the DEEP2 Galaxy Redshift Survey results have
been obtained by \citet{coil2006} at $z\sim1$. The last two clustering
measurements are at similar redshifts to VIPERS.

The bottom panels of Fig.~\ref{fig:r0_gamma} show the stellar mass
dependence of $r_0$ and $\gamma$. We compare our results to data from
SDSS \citep{li2006}, VVDS-Deep \citep{meneux2008}, zCOSMOS
\citep{meneux2009}, NMBS \citep{wake2011}, DEEP2 \citep{mostek2013}
and BzK \citep{lin2012}.  The best-fit clustering lengths from
SDSS-DR2 data in the local Universe \citep{li2006} have been fitted by
\citet{mostek2013}.  The measurements at intermediate redshifts
similar to VIPERS are the VVDS-Deep results obtained at $z\sim0.85$ by
\citet{meneux2008}, the DEEP2 measurements at $z\sim1$ from
\citet{mostek2013}, the clustering measurements by \citet{wake2011},
using NMBS photometric redshifts in the range $0.9<z<1.3$, obtained
with a fixed slope $\gamma=1.6$, and the zCOSMOS measurements by
\citet{meneux2009} at $0.5<z<1$.  The best-fit values of $r_0$ and
$\gamma$ of zCOSMOS have been obtained by fitting the
\citet{meneux2009} data on the scales $r_p<10$\Mpch. All the remaining
measurements refer to higher redshift data. The \citet{wake2011} data
have been taken at the two redshift ranges $1.2<z<1.75$ and
$1.6<z<2.2$. Finally, the BzK $r_0$ values have been estimated from
the angular clustering analysis of star-forming galaxies at $z\sim2$,
for a fixed slope of $\gamma=1.8$.

The general picture that emerges from these studies is that the
clustering correlation length depends significantly on both luminosity
and stellar mass, increasing monotonically from faint and low massive
objects to luminous and massive ones, while the clustering slope
remains fairly constant. Looking specifically to the luminosity
dependence, it is interesting to note that for the intrinsically
brightest samples, $M_{\rm B} > M_{\rm B}^*$, the clustering amplitude
$r_0$ does not evolve significantly from the $z\sim1$ measurements of
VVDS and DEEP2 to the local references of 2dFGRS and SDSS.  The same
behaviour is seen in VIPERS: we find little evolution in the most
luminous bin.  Additionally, the VIPERS data allow us to track $r_0$
with precision over a range of luminosities from $z=0.5-1.1$, showing
that the evolution becomes more significant at fainter luminosities. A
simple qualitative interpretation of these observations is that
low-luminosity galaxies follow the growth of structure approximately,
without any substantial evolution in their bias factor. Conversely,
luminous objects ($M_{\rm B}>M_{\rm B}^*$) need to have a strong
evolution of their bias factor from high to low redshifts, so as to
apparently compensate for the growth of structure and produce the
observed effect.  These considerations are corroborated by the direct
estimate of the bias factor for the different classes of galaxies,
which we discuss in the next section, \S \ref{sec:bias}.

\begin{figure*}
  \includegraphics[width=\textwidth]{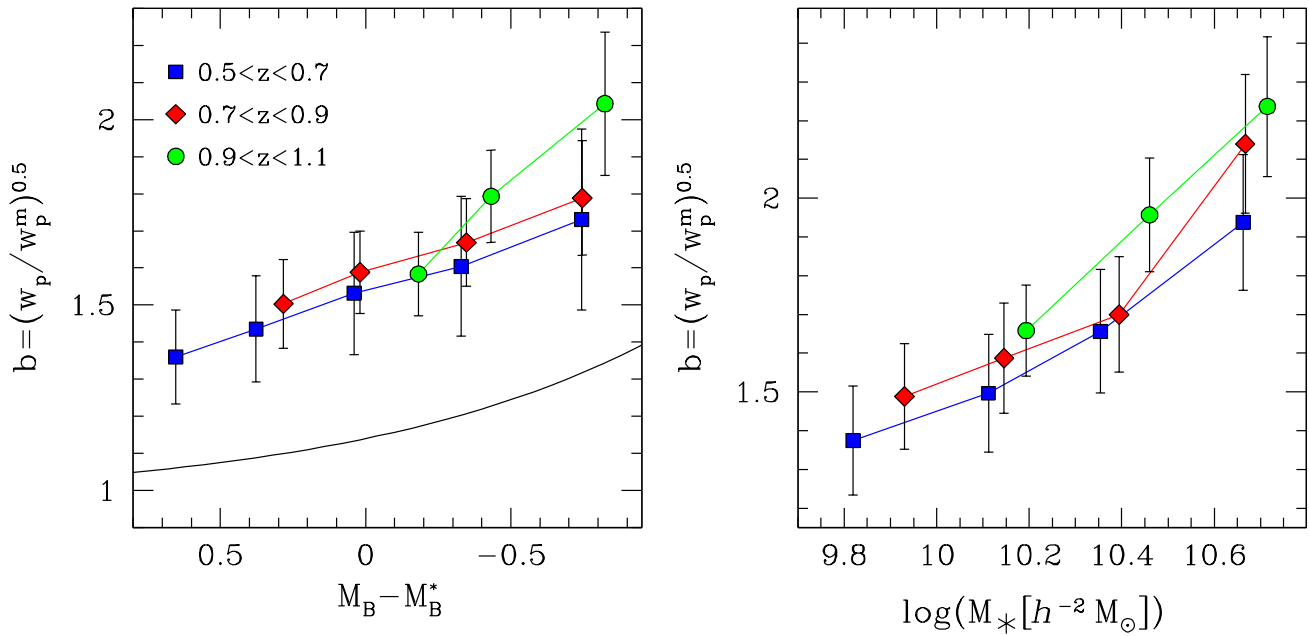}
  \caption{Bias of VIPERS galaxies, averaged over the range $[1-10]$
    \Mpch, as a function of B-band absolute magnitude ({\em left
      panel}) and stellar mass ({\em right panel}). The coloured
    symbols used for VIPERS measurements are the same as in
    Fig.~\ref{fig:r0_gamma}. The black solid line in the left panel
    shows the fitted bias function by \citet{zehavi2011}.}
  \label{fig:bias}
\end{figure*}

On the other hand, when it comes to dependence of clustering on
stellar mass, not all previous measurements are consistent. In
particular, as noted in the bottom left-hand panel of
Fig.~\ref{fig:r0_gamma}, there is a big scatter between previous
measurements, even when they probe similar redshift ranges. Clustering
lengths at $z\sim1$ found in the VVDS-Deep and DEEP2 surveys are
significantly lower not only than the local measurements from the
SDSS-DR7, but also than the values found by NMBS at $0.9<z<2.2$, by
the BzK at $z\sim2$ and by zCOSMOS at $0.5<z<1$.  Our estimates of
$r_0$ are higher than the VVDS-Deep and DEEP2 ones, in the stellar
mass range considered, while they agree fairly well with all the other
data.

The differences are likely to arise from a combination of systematic
effects.  The surveys considered in Fig.~\ref{fig:r0_gamma} are
characterized by different properties, such as the methods for mass
estimation, the cosmological parameters adopted to assess the
distances, and the selection criteria. The last certainly introduce
the most relevant effects, as different surveys can be affected by
stellar mass incompleteness to varying degrees. In particular, the
VIPERS mass-selected samples contain, on average, much brighter
galaxies than VVDS and DEEP2, owing to VIPERS higher flux selection
cut.
 
Moreover, since the power-law model does not describe the data
perfectly, measurements obtained from fitting over different ranges
could produce different values of the fit parameters. This effect is
especially significant for the most massive bins, where the shape of
$w_p(r_p)$ is the furthest from a power law and, thus, most sensitive
to the range, binning, and choice of points used for fitting.
Finally, the impact of sample variance has been estimated with
different techniques, and it might be underestimated in some cases.

The VIPERS clustering slope is almost constant, considering the
estimated errors, around $\gamma\sim1.7-1.8$, with only tentative
evidence of a slight luminosity and redshift evolution.  This agrees
overall with all previous data, except for the relation measured in
VVDS-Deep.  Using the data from the VVDS-Deep Survey,
\citet{pollo2006} found a significant steepening of the correlation
slope from $\gamma\sim1.6$ at $<M_{\rm B}>=19.6$, to $\gamma\sim2.4$
at $<M_{\rm B}>=-21.3$, fitting $w_p(r_p)$ for separation $<10$ \Mpch
in their high-redshift range $0.5<z<1.2$.

To directly compare these results with VVDS data, we measured the
VIPERS 2PCF selecting all galaxies within the same redshift range used
by \citet{pollo2005}, $0.5<z<1.2$. In this way, any possible effect of
the intrinsic evolution in the VVDS broad redshift slice should also
be present in our measurements.  We considered four volume-limited
samples, with the following B-band magnitude thresholds $M_{\rm
  B}(z=1.2)-5\log(h)<-21, -21.25, -21.5, -21.75$. The best-fit values
of $\gamma$ are shown in Fig.~\ref{fig:gamma_MagB} and compared to the
VVDS-Deep measurements in volume-limited samples. To investigate the
scale dependence of the clustering slope, we compared the $\gamma$
values obtained fitting $w_p(r_p)$ in the interval
$0.2<r_p[$\Mpch$]<20$, and considering only small scales, $r_p<1$
\Mpch. The proximity effect caused by the one-pass strategy used in
VIPERS is corrected with the method described in Section \ref{sub:ww}
and Appendix~\ref{appendix}.  We are confident that the correction
recovers the shape of the correlation function on small scales to
within 5\% \citep{delatorre2013b}.

When the 2PCF is fitted on scales up to 20\Mpch, $\gamma$ appears
almost constant at the value $\sim1.7$, as in
Fig.~\ref{fig:gamma_MagB}, while if we fit only the small scales,
$\gamma$ results to be larger and slightly increasing, from
$\gamma\sim1.8$ at $\langle M_{\rm B}-M_{\rm B}^*\rangle\sim-0.3$, to
$\gamma\sim2$ at $\langle M_{\rm B}-M_{\rm B}^*\rangle\sim-0.9$. The
VIPERS clustering slope is still lower than the VVDS one at high
luminosities, though the mismatch is not large (below $2 \sigma$
significance).

In conclusion, we find that the steepening of the slope in VVDS can be
partly attributed to the range of scales used in the fit.  Sample
variance, if not fully recovered from the scatter between the mocks
considered in these analyses, could explain the remaining discrepancy.

\subsection{The galaxy bias}
\label{sec:bias}

\begin{figure*}
  \includegraphics[width=\textwidth]{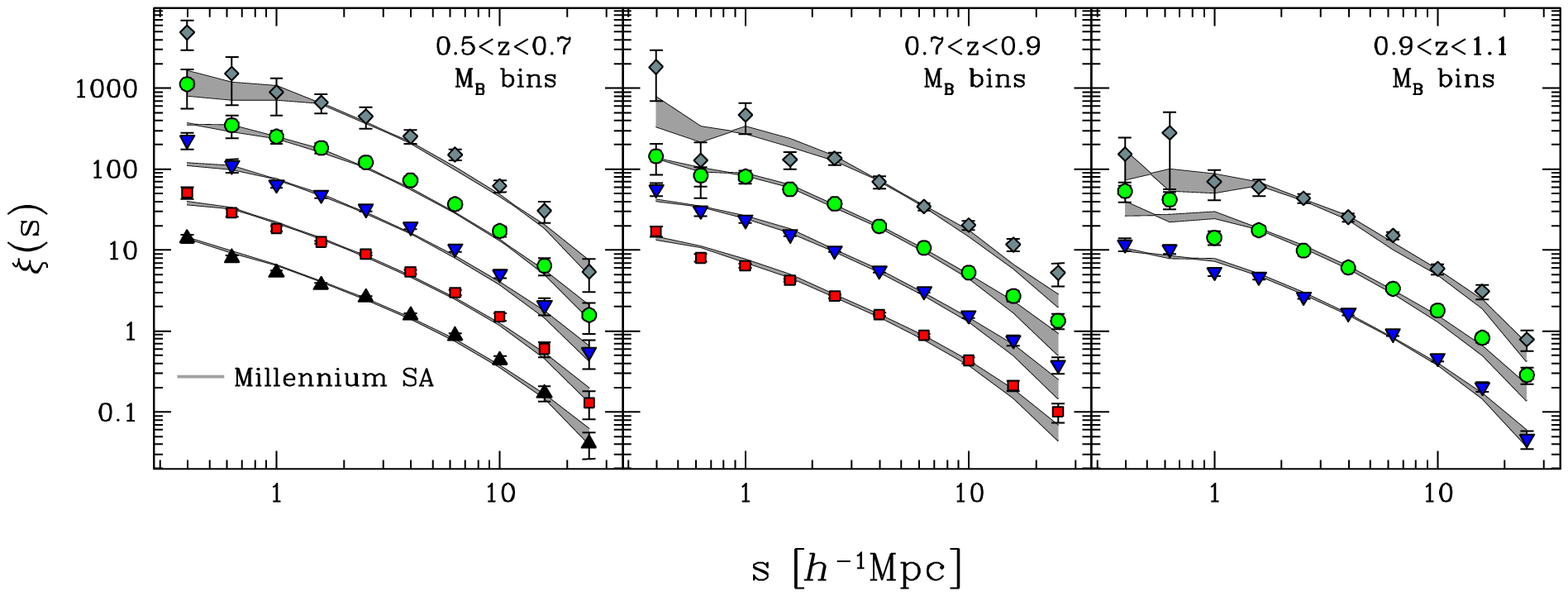}
  \includegraphics[width=\textwidth]{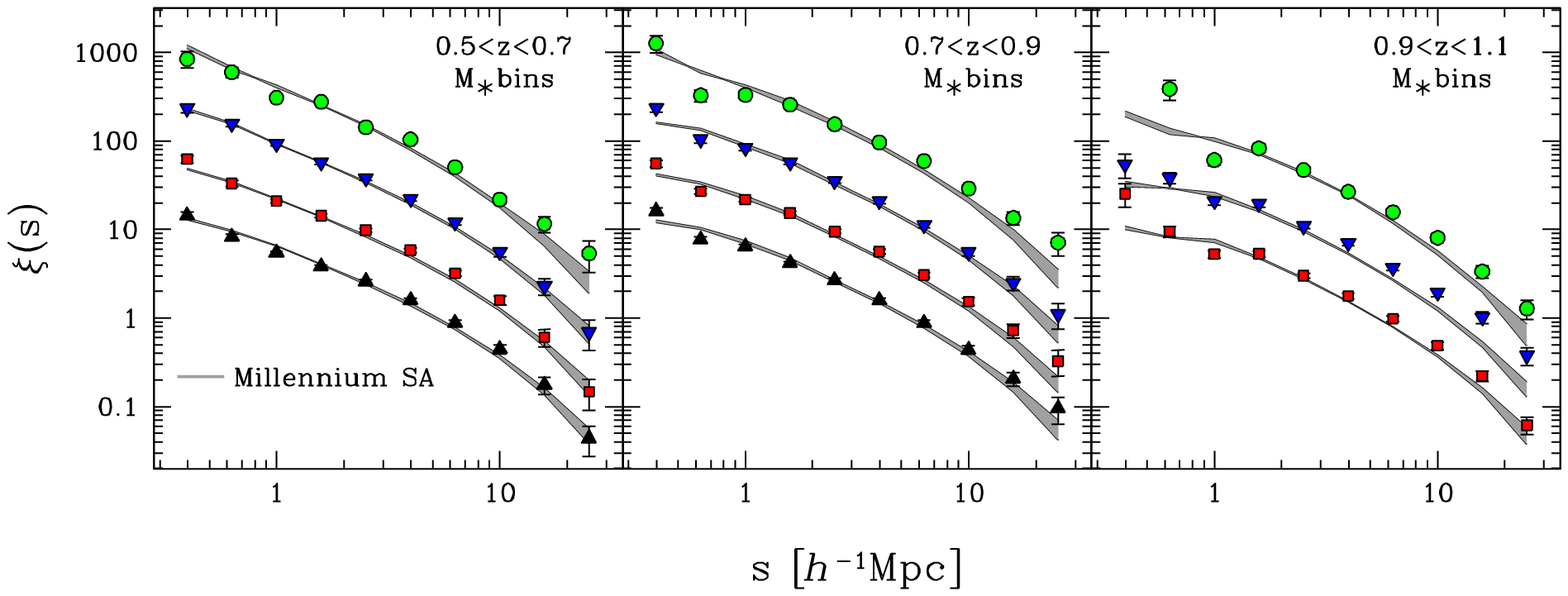}
  \caption{Redshift-space 2PCF of VIPERS galaxies compared to
    theoretical expectations as a function of B-band magnitude ({\em
      upper panels}) and stellar mass ({\em bottom panels}). The
    coloured symbols used for VIPERS measurements are the same as in
    Fig.~\ref{fig:xi_zspace}.  The grey shaded areas show the 2PCF
    measured in two large galaxy mock catalogues, constructed with the
    Munich semi-analytic model. The 2PCFs in different sub-samples are
    offset by $0.5$ dex, starting from the lower luminosity and
    stellar mass samples, for visual clarity.}
  \label{fig:models}
\end{figure*}

In this section, we present our estimates of the large-scale bias of
VIPERS galaxies as a function of luminosity and stellar mass.  We
infer the scale-dependent bias, $b(r_p)$, using
Eq.~\ref{eq:bias}. Then, we average $b(r_p)$ in the range
$1<r_p[$\Mpch$]<10$, where the galaxy bias is fairly constant. To
estimate $b(r_p)$, we assume a flat $\Lambda $CDM model with the
matter density parameter $\Omega_{\rm M}=0.25$, the baryon density
parameter $\Omega_{\rm b}=0.045$, the scalar spectral index $n_s=1$,
and the power spectrum normalization $\sigma_8=0.8$.  A full HOD
formalism would be necessary to properly describe the bias relation
between galaxies and DM on different scales. However, on large scales,
the bias obtained from the ratio of the galaxy and matter correlation
functions is consistent with those obtained using HOD modelling
\citep[see e.g.][]{zehavi2011}.

In Fig.~\ref{fig:bias}, we present the galaxy bias estimated in each
threshold VIPERS sample, as a function of B-band absolute magnitude
(left panel) and stellar mass (right panel). The error bars show the
scatter among the bias values measured in the mock catalogues.  The
values of the bias and its error in each VIPERS sample are reported in
Tables~\ref{tab:table1}--\ref{tab:table4}. The galaxy biasing
estimated with Eq.~\ref{eq:bias} depends on the assumed cosmological
model through both $\sigma_8$ and $\Omega_{\rm M}$.  Its amplitude
scales approximately as $\sigma_8^{-1}$, since $\xi=b^2\xi_m\propto
b^2\sigma_8^2$ on large scales.  The value of $\sigma_8$ is known with
an error of $\sim$2\%, so only just negligible when compared to
uncertainties in our bias parameter ($\sim$4\%).  Uncertainties on
$\Omega_{\rm M}$, which is constrained in the interval
$\sim0.22-0.28$, also have an impact. Varying its input value in that
range induces a $\sim$2\% shift in the estimated bias.  On the other
hand, {\it relative} bias between different galaxy types (e.g. the
ratio between any two bias values reported in the tables), is almost
model-independent.

Figure \ref{fig:bias} provides a compact view of the dependence of
VIPERS clustering on luminosity and stellar mass. More luminous and
massive galaxies appear more strongly biased than fainter and less
massive ones at all redshifts considered.  At fixed luminosity and
stellar mass, the bias function increases with redshift, while the
amplitude of the 2PCF decreases (see Fig.~\ref{fig:r0_gamma}).  In the
left-hand panel of Fig.~\ref{fig:bias}, we show the luminosity
dependence of the galaxy bias in the local Universe, estimated by
\citet{zehavi2011} using the SDSS-DR7 data \citep[see also][for
  similar results]{norberg2001, tegmark2004}.  We note that the
normalization of the local bias is lower than our findings at
intermediate redshifts, as predicted by evolution models \citep[see
  e.g.][]{tegmark1998, kauffmann1999, blanton2000, benson2001,
  pillepich2010}.

As already noted in the previous section, the VIPERS bias in
low-mass-selected samples is biased toward high values compared to
VVDS-Deep and DEEP2 measurements, as a consequence of the $i_{\rm
  AB}<22.5$ selection. This has to be taken into account when
comparing to theoretical predictions; i.e., the same flux selection has
to be applied also in mock catalogues to introduce the same
incompleteness of the VIPERS data.

A more detailed investigation on the scale dependence of the VIPERS
galaxy bias derived with the deprojection method \citep[see
  e.g.][]{marulli2012b} and with a full HOD approach is deferred to a
future work, while the non-linearity of the biasing function is
presented in \citet{diporto2013}.

\subsection{Comparison to theoretical models}
\label{sec:models}

To provide a first theoretical interpretation of our measurements in a
cosmological context, we compared the VIPERS clustering presented in
previous sections with predictions of a standard galaxy formation
model. The result is shown in Fig.~\ref{fig:models}, which compares
the redshift-space 2PCF of VIPERS galaxies, measured in threshold
samples of different luminosities (upper panels) and stellar masses
(lower panels), with the theoretical predictions of the Munich
semi-analytic model \citep{delucia2007}. The VIPERS measurements and
error bars are the same as in Fig.~\ref{fig:xi_zspace}. The
correlations in different sub-samples are offset by $0.5$ dex,
starting from the lower luminosity and stellar mass samples, for
visual clarity.  The Munich semi-analytic model, already introduced in
Section \ref{sub:proxy_mass} to calibrate the corrections for the
proximity effect and stellar mass incompleteness, has been developed
on the outputs of the Millennium run, a large N-body DM simulation of
the $\Lambda $CDM cosmological framework \citep{springel2005}.  The
area of the two galaxy mock catalogues used here is $10x3$ deg$^2$
each, slightly larger than the one of the complete VIPERS survey
($\sim 24$ deg$^2$). Only two independent mock catalogues of this
volume can be extracted from the Millennium simulation.  To estimate
the sample variance, we therefore used our VIPERS HOD mocks, as in the
previous sections. To directly compare our measurements with mock
predictions, we introduced the VIPERS flux limit of $i_{\rm AB}<22.5$
in the mock catalogues.
 
As can be seen, our measurements are compatible with the expectations
of the standard $\Lambda $CDM+semi-analytic framework. Both the
luminosity and stellar mass dependence of VIPERS clustering appear in
overall good agreement with theoretical expectations, though we note
that the most luminous and massive VIPERS galaxies are slightly more
clustered than the semi-analytic ones, especially at high redshift.
We notice that the cosmological parameters of the Millennium
simulation are based on first-year results from the Wilkinson
Microwave Anisotropy Probe (WMAP1). In particular, the value of
$\sigma_8=0.9$ is significantly higher than the most recent estimates.
If the Millennium predictions are properly rescaled to be consistent
with the seven-year WMAP data \citep{guo2013}, galaxies at $z\sim1$
are predicted to be slightly more clustered for
$M_\star\lesssim10^{10.5} M_\odot$, while for lower masses the
predicted clustering is almost coincident with the WMAP1 case. The
differences are nevertheless quite small, given the uncertainties in
current observational data, over the redshift range $0<z<3$
\citep[see][for more details]{guo2013}.  A thorough comparison between
our measurements and theoretical predictions from different galaxy
formation models is beyond the scope of the present paper, and will be
fully pursued in a future work.


\section{Summary and conclusions}
\label{sec:concl}

In this work, we studied how galaxy clustering depends on luminosity
and stellar mass in the redshift range $0.5<z<1.1$, using the
currently largest complete sample of galaxies with redshift measured
in this range. To quantify the effect, we measured the redshift-space
2PCF of VIPERS galaxies in the first data release, PDR-1, and derived
the real-space clustering and bias assuming a power-law form for the
projected correlation function. Errors were estimated using a set of
mock HOD and SHMR catalogues that mimick the characteristics and the
observational selections of the VIPERS dataset.

The main results of this work can be summarized as follows.

\begin{itemize}

\item We confirm with unprecedented precision at $z=0.5-1.1$ a monotonic
  increase in the clustering length $r_0$, both as a function of
  B-band magnitude and stellar mass, in all the three redshift ranges
  considered. In contrast, the clustering slope, $\gamma$, appears
  quite constant when derived in the range $0.2<r_p[$\Mpch$]<20$.

\item The dependence of the clustering length $r_0$ on luminosity is
  stronger at high redshift.  In terms of the evolution of $r_0$ with
  redshift, we find that there is only a weak trend for the highest
  luminosity galaxies, while significant evolution in $r_0$ is found
  in the lower luminosity samples.

\item The VIPERS measurements are generally consistent with previous
  studies, while establishing a bridge from the local references to
  the high-redshift Universe. This is particularly true for the
  dependence of clustering on luminosity. In terms of dependence on
  stellar mass, VIPERS is very consistent with results from surveys
  with comparable magnitude limit, such as zCOSMOS. However, there is
  a clear overall difference at the 2-$\sigma$ level to previous
  measurements at $z\sim1$ obtained from surveys with fainter flux
  limits, such as VVDS-Deep and DEEP2. A possible interpretation is
  that this is the result of two combined effects.  First, VVDS-Deep
  and DEEP2 are about two magnitudes deeper than VIPERS; for this
  reason, a mass-selected sample is significantly more complete in
  mass than in the case of VIPERS or zCOSMOS, including a larger
  population of low-mass, plausibly less clustered galaxies.  This
  would explain the lower value of $r_0$ observed in the VVDS-Deep and
  DEEP2 samples with $M_\star\lesssim10^{11} M_\odot$, that is, in the
  range of masses in which we know surveys like VIPERS or zCOSMOS are
  incomplete (see Fig.~\ref{fig:sub_samples}).  However, this should
  not be the case above this value, and the correlation length
  measured in the fainter surveys should rapidly converge to the trend
  and amplitude indicated by VIPERS and the other ``bright'' surveys.
  In fact, this is not observed in Fig.~\ref{fig:r0_gamma}.  At least
  part of this could be the result of a different effect, that is, the
  much smaller volume of the deeper surveys compared to VIPERS or
  zCOSMOS.  This is particularly severe for the VVDS-Deep sample,
  which is only 0.5 deg$^2$ in area, against the $\sim 11$ deg$^2$
  effectively covered at the current stage by VIPERS, out of a
  footprint of $\sim 15$ deg$^2$.  This factor of $\sim 20$ between
  the sampled volumes for a given redshift range implies that the
  massive tail, populated by the most strongly clustered objects, is
  largely undersampled in the fainter surveys, compared to VIPERS, so
  biasing the measured clustering of the most massive objects toward
  lower values.  This should be less severe for the DEEP2 survey,
  which is $\sim 4$ deg$^2$ in area, thus larger than zCOSMOS. Still,
  at $M_\star\sim10^{11} M_\odot$ the DEEP2 values of $r_0$ remain
  significantly lower than the others.

\item We provide an estimate of galaxy bias, averaged over the range
  $[1-10]$ \Mpch, as a function of luminosity, stellar mass, and
  redshift, assuming a standard flat $\Lambda $CDM framework.  The
  results are consistent with the predictions of hierarchical galaxy
  formation, although we will devote future work to testing evolution
  models in detail.

\end{itemize}

The measurements of the 2PCF presented here represent a considerable
advance over past surveys at $z\sim1$.  The large volume of VIPERS
ensures that sample variance is subdominant.  Furthermore, we modelled
the measurement covariance utilizing mock surveys that account for
both astrophysical and observational effects in detail.  The results
constitute new, stringent constraints on the galaxy 2PCF over a broad
range of luminosities and stellar mass to be used as references
against which to test models of cosmology and galaxy formation at
redshifts $0.5-1.1$.


\section*{Acknowledgments}

We warmly thank Simon D. M. White for helpful suggestions.  We
acknowledge the crucial contribution of the ESO staff for the
management of service observations. In particular, we are deeply
grateful to M. Hilker for his constant help and support of this
programme. Italian participation in VIPERS has been funded by INAF
through PRIN 2008 and 2010 programmes. LG acknowledges support of the
European Research Council through the Darklight ERC Advanced Research
Grant (\# 291521). OLF acknowledges support of the European Research
Council through the EARLY ERC Advanced Research Grant (\#
268107). Polish participants have been supported by the Polish
Ministry of Science (grant N N203 51 29 38), the Polish-Swiss Astro
Project (co-financed by a grant from Switzerland, through the Swiss
Contribution to the enlarged European Union), the European Associated
Laboratory Astrophysics Poland-France HECOLS, and a Japan Society for
the Promotion of Science (JSPS) Postdoctoral Fellowship for Foreign
Researchers (P11802). GDL acknowledges financial support from the
European Research Council under the European Community's Seventh
Framework Programme (FP7/2007-2013)/ERC grant agreement n. 202781. WJP
and RT acknowledge financial support from the European Research
Council under the European Community's Seventh Framework Programme
(FP7/2007-2013)/ERC grant agreement n. 202686. WJP is also grateful
for support from the UK Science and Technology Facilities Council
through the grant ST/I001204/1. EB, FM and LM acknowledge the support
from grants ASI-INAF I/023/12/0 and PRIN MIUR 2010-2011. CM is
grateful for support from specific project funding of the {\it
  Institut Universitaire de France} and the LABEX OCEVU. We would also
like to thank the anonymous referee for useful comments and
suggestions.

\include{tables}

\bibliography{bib}


\appendix
\section{Correcting for stellar mass incompleteness and proximity effect} 
\label{appendix}

\begin{figure}
  \includegraphics[width=0.49\textwidth]{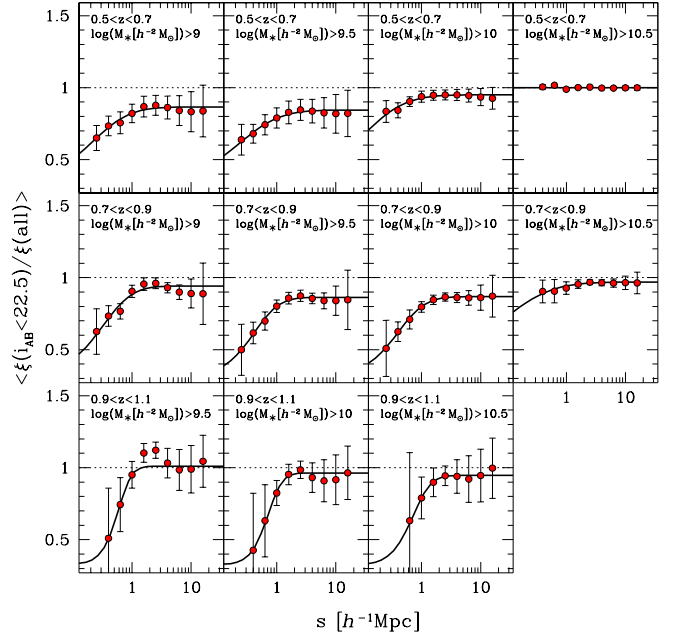}
  \caption{Ratio between the redshift-space 2PCF as a function of
    mass, measured in flux-limited and complete semi-analytic
    mocks. The red dots show the clustering ratio averaged over the
    ten independent mock catalogues, while the error bars are the rms
    scatter. The black lines show the best-fit error function used to
    smooth these ratios.}
  \label{fig:mass_incompl}
\end{figure}

A possible method for correcting the 2PCF measurements for stellar
mass incompleteness is described in Section \ref{sub:proxy_mass}.  We
provide here more details on this procedure and quantify the impact on
our measurements. The method consists in measuring the 2PCF in two
sets of mock galaxy catalogues, one set complete in stellar masses,
the other one with the same VIPERS $i_{AB}<22.5$ flux cut. We use mock
catalogues constructed with the Munich semi-analytic galaxy formation
model \citep{delucia2007}, on top of the DM halo trees from the
Millennium simulation \citep{springel2005}. The {\small MoMaF} package
has been used to produce the light-cones from which the VIPERS mock
samples are extracted \citep{blaizot2005}.  These catalogues are
complete in stellar masses down to $\sim10^{8}M_\odot$.  Figure
\ref{fig:mass_incompl} shows the mean ratio between the 2PCF in
flux-limited and complete mocks, averaged over the ten independent
catalogues, while the error bars are the rms scatter. The results
refer to different redshift and stellar mass limits, as explicitly
indicated by the labels.  According to the Munich model, the mass
incompleteness introduces a clear scale-dependent reduction of
clustering, mainly significant on scales $\lesssim 1$ \Mpch. The
best-fit error functions used to smooth these ratios represent the
correction that we actually use to obtain the values of $r_0^C$,
$\gamma^C$ and bias$^C$ reported in Tables
~(\ref{tab:table3}--\ref{tab:table4}).  Since the mock red and faint
galaxies are significantly more clustered than the observed ones, this
model-dependent correction should be taken with caution.

\begin{figure}
  \includegraphics[width=0.49\textwidth]{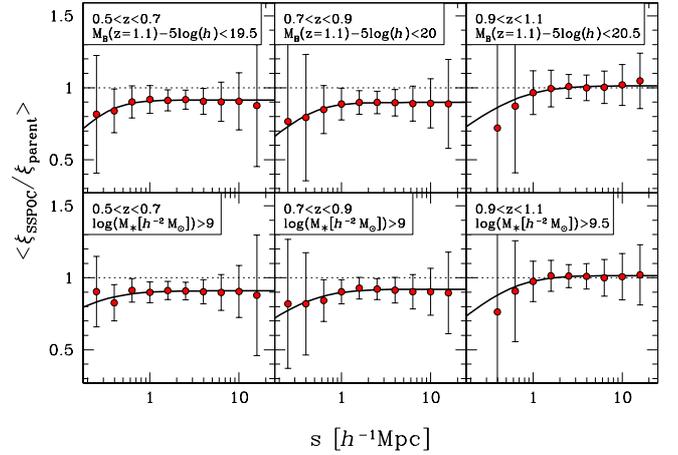}
  \caption{Ratio between the redshift-space 2PCF measured in mock
    catalogues with and without the slit mask target selection
    algorithm applied. Symbols are the same as in
    Fig.~\ref{fig:mass_incompl}.}
  \label{fig:SSPOC}
\end{figure}

As discussed in Section \ref{sub:ww}, owing to our multi-object slit
masks, it is not possible to observe galaxies that are too close to
each other. This proximity effect introduces a further scale-dependent
suppression in the 2PCF that we have to correct. We adopt the same
strategy described above to correct for mass incompleteness; that is,
we measure the ratio between the redshift-space 2PCF in mock
catalogues with and without the slit mask target selection algorithm
applied \citep[also see][]{coil2008}. As before, the mock galaxy
catalogues used here are constructed with the Munich semi-analytic
galaxy formation model.  Figure \ref{fig:SSPOC} shows the results of
this method. This correction is almost independent of luminosity and
stellar mass, similar to the results of \citet{coil2006}. Thus, we
decided to use the correction calibrated in our largest galaxy sample
to correct all our clustering measurements. We notice that the
systematic clustering suppression due to the proximity effect becomes
less severe going to high redshift. More details can be found in
\citet{delatorre2013b}.

\clearpage

\end{document}

%% file: authors.tex
\author{
F.~Marulli\inst{\ref{17},\ref{9},\ref{18}}
\and M.~Bolzonella\inst{\ref{9}}  
\and E.~Branchini\inst{\ref{10},\ref{28},\ref{29}}
\and I.~Davidzon\inst{\ref{17},\ref{9}}
\and S.~de la Torre\inst{\ref{14}}
\and B.~R.~Granett\inst{\ref{2}}
\and L.~Guzzo\inst{\ref{2},\ref{27}}
\and A.~Iovino\inst{\ref{2}}
\and L.~Moscardini\inst{\ref{17},\ref{9},\ref{18}}
\and A.~Pollo\inst{\ref{22},\ref{23}}
\and U.~Abbas\inst{\ref{5}}
\and C.~Adami\inst{\ref{4}}
\and S.~Arnouts\inst{\ref{4},\ref{6}}
\and J.~Bel\inst{\ref{7}}        
\and D.~Bottini\inst{\ref{3}}
\and A.~Cappi\inst{\ref{9},\ref{30}}
\and J.~Coupon\inst{\ref{12}}
\and O.~Cucciati\inst{\ref{9}}     
\and G.~De Lucia\inst{\ref{13}}
\and A.~Fritz\inst{\ref{3}}
\and P.~Franzetti\inst{\ref{3}}
\and M.~Fumana\inst{\ref{3}}
\and B.~Garilli\inst{\ref{4},\ref{3}}    
\and O.~Ilbert\inst{\ref{4}}
\and J.~Krywult\inst{\ref{15}}
\and V.~Le Brun\inst{\ref{4}}
\and O.~Le F\`evre\inst{\ref{4}}
\and D.~Maccagni\inst{\ref{3}}
\and K.~Ma{\l}ek\inst{\ref{16}}
\and H.~J.~McCracken\inst{\ref{19}}
\and L.~Paioro\inst{\ref{3}}
\and M.~Polletta\inst{\ref{3}}
\and H.~Schlagenhaufer\inst{\ref{24},\ref{20}}
\and M.~Scodeggio\inst{\ref{3}} 
\and L.~A.~M.~Tasca\inst{\ref{4}}
\and R.~Tojeiro\inst{\ref{11}}
\and D.~Vergani\inst{\ref{25}}
\and A.~Zanichelli\inst{\ref{26}}
\and A.~Burden\inst{\ref{11}}
\and C.~Di Porto\inst{\ref{9}}
\and A.~Marchetti\inst{\ref{2},\ref{1}} 
\and C.~Marinoni\inst{\ref{7}}
\and Y.~Mellier\inst{\ref{19}}
\and R.~C.~Nichol\inst{\ref{11}}
\and J.~A.~Peacock\inst{\ref{14}}
\and W.~J.~Percival\inst{\ref{11}}
\and S.~Phleps\inst{\ref{20}}
\and M.~Wolk\inst{\ref{19}}
\and G.~Zamorani\inst{\ref{9}}
}
\offprints{F.Marulli \\ \email{federico.marulli3@unibo.it}}
\institute{
Dipartimento di Fisica e Astronomia - Universit\`{a} di Bologna, viale Berti Pichat 6/2, I-40127 Bologna, Italy \label{17}
\and INAF - Osservatorio Astronomico di Bologna, via Ranzani 1, I-40127 Bologna, Italy \label{9}
\and INFN - Sezione di Bologna, viale Berti Pichat 6/2, I-40127 Bologna, Italy \label{18}
\and Dipartimento di Matematica e Fisica, Universit\`{a} degli Studi Roma Tre, via della Vasca Navale 84, I-00146 Roma, Italy \label{10}
\and INFN - Sezione di Roma Tre, via della Vasca Navale 84, I-00146 Roma, Italy \label{28}
\and INAF - Osservatorio Astronomico di Roma, via Frascati 33, I-00040 Monte Porzio Catone (RM), Italy \label{29}
\and SUPA - Institute for Astronomy, University of Edinburgh, Royal Observatory, Blackford Hill, Edinburgh, EH9 3HJ, UK \label{14}
\and INAF - Osservatorio Astronomico di Brera, Via Brera 28, 20122 Milano, via E. Bianchi 46, I-23807 Merate, Italy \label{2}
\and Dipartimento di Fisica, Universit\`a di Milano-Bicocca, P.zza della Scienza 3, I-20126 Milano, Italy \label{27}
\and Astronomical Observatory of the Jagiellonian University, Orla 171, 30-001 Cracow, Poland \label{22}
\and National Centre for Nuclear Research, ul. Hoza 69, 00-681 Warszawa, Poland \label{23}
\and INAF - Osservatorio Astronomico di Torino, I-10025 Pino Torinese, Italy \label{5}
\and Aix Marseille Universit\'e, CNRS, LAM (Laboratoire d'Astrophysique de Marseille) UMR 7326, 13388, Marseille, France  \label{4}
\and Canada-France-Hawaii Telescope, 65--1238 Mamalahoa Highway, Kamuela, HI 96743, USA \label{6}
\and Centre de Physique Th\'eorique, UMR 6207 CNRS-Universit\'e de Provence, Case 907, F-13288 Marseille, France \label{7}
\and INAF - Istituto di Astrofisica Spaziale e Fisica Cosmica Milano, via Bassini 15, I-20133 Milano, Italy\label{3}
\and Laboratoire Lagrange, UMR7293, Universit\'e de Nice Sophia-Antipolis,  CNRS, Observatoire de la C\^ote d'Azur, 06300 Nice, France \label{30}
\and Institute of Astronomy and Astrophysics, Academia Sinica, P.O. Box 23-141, Taipei 10617, Taiwan\label{12}
\and INAF - Osservatorio Astronomico di Trieste, via G. B. Tiepolo 11, I-34143 Trieste, Italy \label{13}
\and Institute of Physics, Jan Kochanowski University, ul. Swietokrzyska 15, 25-406 Kielce, Poland \label{15}
\and Department of Particle and Astrophysical Science, Nagoya University, Furo-cho, Chikusa-ku, 464-8602 Nagoya, Japan \label{16}
\and Institute d'Astrophysique de Paris, UMR7095 CNRS, Universit\'{e} Pierre et Marie Curie, 98 bis Boulevard Arago, 75014 Paris, France \label{19}
\and Universit\"{a}tssternwarte M\"{u}nchen, Ludwig-Maximillians Universit\"{a}t, Scheinerstr. 1, D-81679 M\"{u}nchen, Germany \label{24}
\and Max-Planck-Institut f\"{u}r Extraterrestrische Physik, D-84571 Garching b. M\"{u}nchen, Germany \label{20}
\and Institute of Cosmology and Gravitation, Dennis Sciama Building, University of Portsmouth, Burnaby Road, Portsmouth, PO1 3FX, UK \label{11}
\and INAF - Istituto di Astrofisica Spaziale e Fisica Cosmica Bologna, via Gobetti 101, I-40129 Bologna, Italy \label{25}
\and INAF - Istituto di Radioastronomia, via Gobetti 101, I-40129 Bologna, Italy \label{26}
\and Universit\`{a} degli Studi di Milano, via G. Celoria 16, I-20130 Milano, Italy \label{1}
}

%% file: tables.tex
\begin{table*}
  \begin{center}
    \caption[]{Properties of the selected VIPERS sub-samples in
      threshold luminosity bins. The best-fit values of the clustering
      normalization, $r_0$, and slope, $\gamma$, have been obtained by
      fitting $w_p(r_p)$ in $0.2<r_p[$\Mpch$]<20$. The errors are from
      the scatter among the HOD VIPERS mocks. The bias values assume a
      flat $\Lambda $CDM model with $\Omega_{\rm M}=0.25$ and
      $\sigma_8=0.8$.}
    \begin{tabular}{cccccccc}
      \hline
      \hline
      redshift & median & magnitude range & median magnitude & $N_{gal}$ & $r_0$ & $\gamma$ & bias \\
      range & redshift & $M_B(z=1.1)-5\log(h)$ & $M_B-5\log(h)$ & & [\Mpch] & & \\
      \hline
      \hline
      $[0.5,0.7]$ & $0.62$ & $<-19.5$ & $-19.87$ & $17473$ & $4.45\pm0.20$ & $1.67\pm0.05$ & $1.36\pm0.13$ \\
      $[0.5,0.7]$ & $0.62$ & $<-20.0$ & $-20.15$ & $12432$ & $4.81\pm0.21$ & $1.69\pm0.05$ & $1.44\pm0.14$ \\
      $[0.5,0.7]$ & $0.62$ & $<-20.5$ & $-20.49$ & $7472$ & $5.22\pm0.29$ & $1.73\pm0.06$ & $1.53\pm0.17$ \\
      $[0.5,0.7]$ & $0.62$ & $<-21.0$ & $-20.86$ & $3599$ & $5.58\pm0.39$ & $1.81\pm0.08$ & $1.60\pm0.19$ \\
      $[0.5,0.7]$ & $0.62$ & $<-21.5$ & $-21.28$ & $1236$ & $6.24\pm0.56$ & $1.79\pm0.14$ & $1.73\pm0.24$ \\
      \hline
      $[0.7,0.9]$ & $0.79$ & $<-20.0$ & $-20.41$ & $14442$ & $4.56\pm0.22$ & $1.65\pm0.04$ & $1.50\pm0.12$ \\
      $[0.7,0.9]$ & $0.80$ & $<-20.5$ & $-20.68$ & $9469$ & $4.95\pm0.23$ & $1.67\pm0.04$ & $1.59\pm0.11$ \\
      $[0.7,0.9]$ & $0.80$ & $<-21.0$ & $-21.05$ & $4605$ & $5.32\pm0.30$ & $1.70\pm0.05$ & $1.67\pm0.12$ \\
      $[0.7,0.9]$ & $0.80$ & $<-21.5$ & $-21.45$ & $1619$ & $5.95\pm0.39$ & $1.72\pm0.10$ & $1.79\pm0.15$ \\
      \hline
      $[0.9,1.1]$ & $0.97$ & $<-20.5$ & $-21.00$ & $5207$ & $4.29\pm0.19$ & $1.63\pm0.04$ & $1.58\pm0.11$ \\
      $[0.9,1.1]$ & $0.98$ & $<-21.0$ & $-21.25$ & $3477$ & $5.08\pm0.26$ & $1.64\pm0.05$ & $1.79\pm0.12$ \\
      $[0.9,1.1]$ & $0.99$ & $<-21.5$ & $-21.65$ & $1409$ & $5.87\pm0.43$ & $1.68\pm0.08$ & $2.04\pm0.19$ \\
      \hline
      \hline
      \label{tab:table1}
    \end{tabular}
  \end{center}
\end{table*}
\begin{table*}
  \begin{center}
    \caption[]{Properties of the selected VIPERS sub-samples in
      binned luminosity bins. Parameters and errors are as in
      Table~\ref{tab:table1}.}
    \begin{tabular}{cccccccc}
      \hline
      \hline
      redshift range & median & magnitude range & median magnitude & $N_{gal}$ & $r_0$ & $\gamma$ & bias \\
      range & redshift & $M_B(z=1.1)-5\log(h)$ & $M_B-5\log(h)$ & & [\Mpch] & & \\
      \hline
      \hline
      $[0.5,0.7]$ & $0.62$ & [$-20.5,-19.5]$ & $-19.51$ & $10001$ & $3.92\pm0.19$ & $1.65\pm0.05$ & $1.24\pm0.12$ \\
      $[0.5,0.7]$ & $0.62$ & [$-21.0,-20.0]$ & $-19.95$ & $8833$ & $4.58\pm0.21$ & $1.66\pm0.05$ & $1.39\pm0.14$ \\
      $[0.5,0.7]$ & $0.62$ & [$-21.5,-20.5]$ & $-20.40$ & $6236$ & $5.00\pm0.29$ & $1.72\pm0.06$ & $1.47\pm0.16$ \\
      $[0.5,0.7]$ & $0.62$ & [$-22.0,-21.0]$ & $-20.83$ & $3334$ & $5.35\pm0.36$ & $1.78\pm0.08$ & $1.56\pm0.18$ \\
      $[0.5,0.7]$ & $0.62$ & [$-22.5,-21.5]$ & $-21.27$ & $1207$ & $6.29\pm0.59$ & $1.80\pm0.13$ & $1.75\pm0.24$ \\
      \hline
      $[0.7,0.9]$ & $0.78$ & [$-21.0,-20.0]$ & $-20.19$ & $9837$ & $4.08\pm0.21$ & $1.61\pm0.05$ & $1.40\pm0.12$ \\
      $[0.7,0.9]$ & $0.80$ & [$-21.5,-20.5]$ & $-20.59$ & $7850$ & $4.71\pm0.24$ & $1.67\pm0.04$ & $1.53\pm0.11$ \\
      $[0.7,0.9]$ & $0.80$ & [$-22.0,-21.0]$ & $-21.01$ & $4246$ & $5.20\pm0.31$ & $1.68\pm0.05$ & $1.66\pm0.12$ \\
      $[0.7,0.9]$ & $0.80$ & [$-22.5,-21.5]$ & $-21.45$ & $1586$ & $5.96\pm0.40$ & $1.71\pm0.10$ & $1.80\pm0.16$ \\
      \hline
      $[0.9,1.1]$ & $0.96$ & [$-21.5,-20.5]$ & $-20.86$ & $3895$ & $3.63\pm0.15$ & $1.64\pm0.04$ & $1.36\pm0.10$ \\
      $[0.9,1.1]$ & $0.98$ & [$-22.0,-21.0]$ & $-21.21$ & $2777$ & $4.90\pm0.29$ & $1.65\pm0.04$ & $1.74\pm0.12$ \\
      $[0.9,1.1]$ & $0.99$ & [$-22.5,-21.5]$ & $-21.64$ & $1091$ & $5.98\pm0.43$ & $1.70\pm0.08$ & $2.03\pm0.19$ \\
      \hline
      \hline
      \label{tab:table2}
    \end{tabular}
  \end{center}
\end{table*}
\begin{table*}
  \begin{center}
    \caption[]{Properties of the selected VIPERS sub-samples in
      threshold stellar mass bins.  The quantities $r_0^{C}$,
      $\gamma^{C}$, and bias$^{C}$ are the clustering normalization,
      slope, and linear bias corrected for the stellar mass
      incompleteness, respectively. The other parameters and errors
      are the same as in Table~\ref{tab:table1}.}
    \begin{tabular}{ccccccccccc}
      \hline
      \hline
      redshift & median & stellar mass range & median stellar mass & $N_{gal}$ & $r_0$ & $\gamma$ & bias & $r_0^{C}$ & $\gamma^{C}$ & bias$^{C}$ \\
      range & redshift & $[h^{-2}\,M_\odot]$ & $[h^{-2}\,M_\odot]$ & & [\Mpch] & & \\
      \hline
      \hline
      $[0.5,0.7]$ & $0.61$ & $>9.0$ & $9.82$ & $17100$ & $4.55\pm0.22$ & $1.69\pm0.05$ & $1.38\pm0.14$ & $4.99$ & $1.70$ & $1.48$ \\
      $[0.5,0.7]$ & $0.62$ & $>9.5$ & $10.11$ & $11567$ & $5.14\pm0.25$ & $1.76\pm0.05$ & $1.50\pm0.15$ & $5.70$ & $1.77$ & $1.63$ \\
      $[0.5,0.7]$ & $0.62$ & $>10.0$ & $10.35$ & $6880$ & $5.84\pm0.28$ & $1.83\pm0.05$ & $1.66\pm0.16$ & $6.01$ & $1.83$ & $1.70$ \\
      $[0.5,0.7]$ & $0.62$ & $>10.5$ & $10.66$ & $2151$ & $6.97\pm0.37$ & $1.86\pm0.06$ & $1.94\pm0.17$ & $6.97$ & $1.86$ & $1.94$ \\
      \hline
      $[0.7,0.9]$ & $0.78$ & $>9.0$ & $9.93$ & $15019$ & $4.53\pm0.25$ & $1.66\pm0.04$ & $1.49\pm0.14$ & $4.73$ & $1.67$ & $1.54$ \\
      $[0.7,0.9]$ & $0.79$ & $>9.5$ & $10.15$ & $11345$ & $4.92\pm0.26$ & $1.67\pm0.04$ & $1.59\pm0.14$ & $5.41$ & $1.69$ & $1.71$ \\
      $[0.7,0.9]$ & $0.79$ & $>10.0$ & $10.39$ & $6884$ & $5.37\pm0.29$ & $1.72\pm0.04$ & $1.70\pm0.15$ & $5.87$ & $1.73$ & $1.83$ \\
      $[0.7,0.9]$ & $0.79$ & $>10.5$ & $10.67$ & $2498$ & $7.08\pm0.38$ & $1.73\pm0.05$ & $2.14\pm0.18$ & $7.22$ & $1.73$ & $2.18$ \\
      \hline
      $[0.9,1.1]$ & $0.97$ & $>9.5$ & $10.19$ & $4558$ & $4.48\pm0.21$ & $1.60\pm0.04$ & $1.66\pm0.12$ & $4.51$ & $1.62$ & $1.65$ \\
      $[0.9,1.1]$ & $0.97$ & $>10.0$ & $10.46$ & $2857$ & $5.57\pm0.27$ & $1.62\pm0.04$ & $1.96\pm0.15$ & $5.78$ & $1.65$ & $2.00$ \\
      $[0.9,1.1]$ & $0.97$ & $>10.5$ & $10.71$ & $1281$ & $6.52\pm0.35$ & $1.74\pm0.05$ & $2.24\pm0.18$ & $6.78$ & $1.76$ & $2.30$ \\
      \hline
      \hline
      \label{tab:table3}
    \end{tabular}
  \end{center}
\end{table*}
\begin{table*}
  \begin{center}
    \caption[]{Properties of the selected VIPERS sub-samples in
      binned stellar mass bins. Parameters and errors are as in
      Table~\ref{tab:table3}.}
    \begin{tabular}{ccccccccccc}
      \hline
      \hline
      redshift & median & stellar mass range & median stellar mass & $N_{gal}$ & $r_0$ & $\gamma$ & bias & $r_0^{C}$ & $\gamma^{C}$ & bias$^{C}$  \\
      range & redshift & $[h^{-2}\,M_\odot]$ & $[h^{-2}\,M_\odot]$ & & [\Mpch] & & \\
      \hline
      \hline
      $[0.5,0.7]$ & $0.61$ & $[9.5,10.5]$ & $10.00$ & $9416$ & $4.75\pm0.26$ & $1.71\pm0.05$ & $1.42\pm0.16$ & $5.59$ & $1.73$ & $1.62$ \\
      $[0.5,0.7]$ & $0.62$ & $[10.0,11.0]$ & $10.35$ & $6705$ & $5.69\pm0.28$ & $1.83\pm0.05$ & $1.61\pm0.16$ & $5.87$ & $1.84$ & $1.66$ \\
      $[0.5,0.7]$ & $0.62$ & $[10.5,11.5]$ & $10.66$ & $2150$ & $6.91\pm0.36$ & $1.87\pm0.06$ & $1.92\pm0.17$ & $6.93$ & $1.87$ & $1.93$ \\
      \hline
      $[0.7,0.9]$ & $0.79$ & $[9.5,10.5]$ & $10.00$ & $8848$ & $4.37\pm0.27$ & $1.62\pm0.05$ & $1.47\pm0.16$ & $5.41$ & $1.63$ & $1.74$ \\
      $[0.7,0.9]$ & $0.79$ & $[10.0,11.0]$ & $10.39$ & $6729$ & $5.28\pm0.28$ & $1.68\pm0.04$ & $1.69\pm0.15$ & $5.84$ & $1.70$ & $1.83$ \\
      $[0.7,0.9]$ & $0.79$ & $[10.5,11.5]$ & $10.67$ & $2497$ & $7.04\pm0.37$ & $1.73\pm0.05$ & $2.13\pm0.18$ & $7.20$ & $1.73$ & $2.18$ \\
      \hline
      $[0.9,1.1]$ & $0.96$ & $[9.5,10.5]$ & $9.98$ & $3277$ & $4.23\pm0.28$ & $1.64\pm0.06$ & $1.52\pm0.13$ & $4.57$ & $1.64$ & $1.61$ \\
      $[0.9,1.1]$ & $0.97$ & $[10.0,11.0]$ & $10.45$ & $2741$ & $5.49\pm0.27$ & $1.61\pm0.04$ & $1.93\pm0.15$ & $5.81$ & $1.64$ & $2.01$ \\
      \hline
      \hline
      \label{tab:table4}
    \end{tabular}
  \end{center}
\end{table*}